\documentclass[letter,12pt]{article}
\usepackage{amsmath}
\usepackage{graphicx,psfrag,epsf}
\usepackage{enumerate}

\usepackage{amssymb}
\usepackage{amsfonts}
\usepackage{mathtools}
\usepackage{bm}
\usepackage{color,soul}
\usepackage[final]{pdfpages}
\usepackage{subfigure}

\usepackage{url}
\usepackage[round,authoryear]{natbib}
\newtheorem{theorem}{Theorem}[section]
\newtheorem{lemma}{Lemma}

\newtheorem{remark}{Remark}

\usepackage{float}
\usepackage[linesnumbered, ruled]{algorithm2e}
\SetKwRepeat{Do}{do}{while}

\usepackage{longtable}
\newcommand{\blind}{1}

\newcommand\norm[1]{\left\lVert#1\right\rVert}
\usepackage{makecell}

\usepackage{enumitem} % for '\newlist' and '\setlist' macros
\newlist{condenum}{enumerate}{1} % 'condenum': a new, enumerate-like list env.
\setlist[condenum]{label=(C\arabic*), 
	ref=(C\arabic*), wide}

\newcommand{\bC}{\boldsymbol{C}}

\newcommand{\bG}{\boldsymbol{G}}
\newcommand{\bg}{\boldsymbol{g}}
\newcommand{\bH}{\boldsymbol{H}}

\newcommand{\bI}{\boldsymbol{\mathrm{I}}}

\newcommand{\bbeta}{\boldsymbol{\beta}}

\newcommand{\mD}{\mathcal{D}}
\newcommand{\mM}{\mathcal{M}}
\newcommand{\mA}{\mathcal{A}}
\newcommand{\tbeta}{\bm{\tilde{\beta}}}

\author{Lan Luo and Peter X.-K. Song}

\begin{document}

\def\spacingset#1{\renewcommand{\baselinestretch}%
	{#1}\small\normalsize} \spacingset{1}

%%%%%%%%%%%%%%%%%%%%%%%%%%%%%%%%%%%%%%%%%%%%%%%%%%%%%%%%%%%%%%%%%%%%%%%%%%%%%%

\if1\blind
{
	\title{\bf Real-Time Regression Analysis of Streaming Clustered Data With Possible Abnormal Data Batches}
	\author{Lan Luo\thanks{Department of Statistics and Actuarial Science, University of Iowa, USA}, \hspace{.2 cm}
		%	The authors gratefully acknowledge \textit{please remember to list all relevant funding sources in the unblinded version}}\hspace{.2cm}\\
	%	Department of Biostatistics, University of Michigan\\
	Ling Zhou\thanks{Center of Statistical Research and School of Statistics, Southwestern University of Finance and Economics, China}, \hspace{.2cm}
		and %\\
		Peter X.-K. Song \thanks{
		Department of Biostatistics, University of Michigan, USA}
	}
	\maketitle
} \fi

\if0\blind
{
	\bigskip
	\bigskip
	\bigskip
	\begin{center}
		{\bf Real-Time Regression Analysis of Streaming Clustered Data With or Without Abnormal Data Batches}
	\end{center}
	\medskip
} \fi

\bigskip
\begin{abstract}
This paper develops an incremental learning algorithm based on quadratic inference function (QIF) to analyze streaming datasets with correlated outcomes such as longitudinal data and clustered data. We propose a renewable QIF (RenewQIF) method within a paradigm of renewable estimation and incremental inference, in which parameter estimates are recursively renewed with current data and summary statistics of historical data, but with no use of any historical subject-level raw data. We compare our renewable estimation method with both offline QIF and offline generalized estimating equations (GEE) approach that process the entire cumulative subject-level data all together, and show theoretically and numerically that our renewable procedure enjoys statistical and computational efficiency. We also propose an approach to diagnose the homogeneity assumption of regression coefficients via a sequential goodness-of-fit test as a screening procedure on occurrences of abnormal data batches. We implement the proposed methodology by expanding existing Spark's Lambda architecture for the operation of statistical inference and data quality diagnosis. We illustrate the proposed methodology by extensive simulation studies and an analysis of streaming car crash datasets from the National Automotive Sampling System-Crashworthiness Data System (NASS CDS). %The supplementary material is available online.
\end{abstract}

\noindent%
\vfill

\newpage
\spacingset{1.5} % DON'T change the spacing!

\section{Introduction}
\label{sec:intro_QIF}

When a car accident happens, driver and passengers in the same car would be all likely to get injured, and their degrees of injury are correlated within a car.  Here a car is a sample unit which, in general, is referred to as a cluster. The National Automotive Sampling System-Crashworthiness Data System (NASS CDS) is a publicly accessible source of streaming datasets containing car accident information in the USA. Other examples of such streaming correlated data include cohorts of patients sequentially assembled at different clinical centers to periodically update national disease registry databases, where, for example, a family is the sample unit. In this paper, we consider a problem where a series of independent clusters becomes available sequentially over data batches, and arrivals of data batches may be perpetual. Similar to the data of car accidents, each data batch consists of temporally correlated or cluster-correlated outcomes. The primary goal of processing such streaming data is to sequentially update some statistics of interest upon the arrival of a new data batch, in the hope to not only free up space for the storage of massive historical individual-level data, but also to provide real-time inference and decision making. 

With the emergence of streaming data collection techniques,  sequential data analytics have received much attention in the literature to address computational efficiency while preserving essential statistical properties.  Arguably, stochastic gradient descent (SGD) algorithm has bee thus far the most well-known algorithm to analyze streaming data along the lines of stochastic approximations~\citep{Robbins1951}. Unfortunately, most of currently available online learning methods in the SGD paradigm  and its variants,  including online Newton SGD~\citep{Log2007} and quasi-Newton SGD~\citep{SGDQN2009,SchYuGue07}, have focused only on point estimation or prediction, and unfortunately precluded statistical inference.~\citet{Toulis2017} obtained some analytic expressions for the asymptotic variances in the implementation of an SGD algorithm to produce maximum likelihood estimation (MLE) with cross-sectional data, in which statistical inference was absent. Recently, \citet{Fang2019} proposed a perturbation-based resampling method to construct confidence intervals in the framework of the averaged implicit SGD (AI-SGD) estimation proposed by~\citet{Toulis2017}. \citet{Luo2020} demonstrated in simulation studies that 
Fang's resampling method for the AI-SGD may fail to provide desirable coverage probability, and thus possibly leads to misleading inference, especially when the number of regression parameters is large. In the setting of linear models, since the least square estimation (LSE) of the regression parameters has a closed form expression for recursive calculations, it is possible to establish certain sequential updating schemes for both point estimation and related calculation of standard errors; see form example \citet{Luo2020}. In this case the resulting estimates can be exactly the same as those obtained by its offline counterpart~\citep{Stengel1994}. This property has been utilized to conduct real-time regression analysis with the linear model for streaming data by, for example,~\cite{StreamFitter2011}. However, according to~\cite{Luo2020}, this property of equality between online and offline estimates no longer holds beyond the linear model, e.g.  the logistic model for binary outcomes.   

This paper considers an important extension of renewable estimation and incremental inference in the class of generalized linear models (GLMs) developed by~\cite{Luo2020} for cross-sectional data to the the case of streaming data with repeatedly measured responses. This extension is substantial in both methodology and applications. In terms of methodological extensions, it relaxes not only the availability of likelihood functions in the renewable MLE to a general framework of estimating functions, but also the popular assumption of homogeneous marginal models. Model homogeneity refers to the situation where data streams are generated under a common set of parameters over the sequence of data batches. This homogeneity will be violated in the presence of ``abnormal'' data batches -- those being generated under different sets of model parameters from the standard/default ones of primary interest.  In real world applications, practitioners often encounter outlying data batches. In this case, continually updating results without noticing and removing abnormal data batches would lead to invalid statistical inference and misleading conclusions. In deal with this issue of practical importance, we develop a quality control (QC) type of monitoring scheme by the means of abnormal data batch detection and deletion. 

We propose a new online regression methodology along the lines of the generalized estimating equation (GEE) approach proposed by~\citet{GEE1986}, one of the most widely used methods for the analysis of data with correlated outcomes. This quasi-likelihood approach is based only on the first two moments of the correlated data distribution with no need of specifying a parametric joint distribution. Such regression model is  termed as marginal generalized linear model (MGLM) or population-average model in the literature of correlated data analysis \citep[Chapter 5]{Song2007correlated}. In this field, another quasi-likelihood inference method is quadratic inference function (QIF) ~\citep{Qu2000}. QIF has several advantages in comparison to GEE: (i) QIF does not require more model assumptions than GEE; (ii) it provides a goodness-of-fit test for the first moment assumption, i.e. the mean-model specification; (iii) QIF estimator is more efficient than the GEE estimator when the working correlation is misspecified; and (iv) it is more robust with a bounded influence function against large outliers~\citep{QuSong2004}.

Our key methodology contributions include: (i) we propose an online QIF method that allows to perform real-time regression analysis of correlated outcomes under fast recursive estimating equations with little reliance on data storage capacity; (ii) the proposed online estimation and inference, termed as RenewQIF, is asymptotically equivalent to the offline QIF estimator obtained from the full cumulative data, and thus has no loss of statistical power in inference; (iii) our RenewQIF method can be implemented in the existing Spark's Lambda architecture to carry over incremental estimation and inference with desirable statistical and computational efficiency; and (iv) by adding a monitoring layer to the Lambda architecture, our method allows to detect and delete abnormal data batches in the real-time analysis of correlated data.  A direct use of the offline QIF with cumulative data encounters fast-growing demands on hardward capacities over the course of perpetual data streams.

It is worth pointing out that the proposed RenewQIF is indeed different from the offline QIF that works for a single data batch. First, the RenewQIF is built upon a sequential updating paradigm, which is computationally much faster than the offline QIF. This gain in computational speed stems from to our new formulation of online search algorithm that is operated recursively under a different objective function from that of the original offline QIF. As shown in this paper, the RenewQIF and offline QIF estimators are not the same but only stochastically equivalent. Second, unlike most of existing online methods, the proposed RenewQIF provides online statistical inference that, technically, depends on a fast recursive calculation of Godambe information matrix (or the sandwich covariance matrix).  Such work has not been considered in the current literature.  Third, theoretical justifications for large sample properties  of the RenewQIF are different in the case where individual data batch sizes are fixed but the number of data batches tends to infinity.  In contrast, the asymptotic results of the offline QIF do not hold if the sample size is fixed, which corresponds to the case that the number of data batches is only one rather than diverging to infinity.  Thus, the technical treatments in both theoretical arguments and algorithm designs in the RenewQIF are more advanced than those given by~\citet{Qu2000}. Technically, the offline QIF may be regarded as a special case of the RenewQIF.

We also propose an online screening method for a real-time diagnosis of abnormal data batches in the framework of RenewQIF. Most existing online monitoring procedures are based on certain metrics such as a test statistic~\citep{Schart1995,EWMA1991,cusum1954a,SR1963, SR1966,Pollak1991,GLR1971}. In a similar spirit to~\citep{Lai2004}, we establish a fast screening procedure based on Hansen's goodness-of-fit test statistic~\citep{Hansen1982} to identify any abnormal data batch and exclude it from updating results of estimation and inference. 
%recursive updates are preferrable when a new data batch arrives. In general, we need to turn a test statistic into a recursive form for online updating~\citep{Lai2004}. In this paper we allow occurrences of abnormal data batches over clustered data streams. By an ``abnormal" data batch we mean a dataset that is generated with a different set of regression coefficients from those of the model of primary interest. We establish a fast screening procedure based on a goodness-of-fit test statistic to identify any abnormal data batch and exclude it from updating results of estimation and inference.
To implement the RenewQIF in the presence of potential abnormal data batches, we expand the Rho architecture developed by~\citep{Luo2020} in the context of GLMs for cross-sectional data, by adding a new monitoring layer in the Spark's Lambda architecture. This monitoring layer houses a QIF-based testing procedure to check the compatibility of each arriving data batch with the normal data reference. 

Essentially, we aim to develop a new online methodology with the following tasks: (i) to put forward RenewQIF estimation and incremental inference in MGLMs for correlated outcomes; and (ii) to study a QIF-based goodness-of-fit test statistic in the monitoring layer that enables to effectively detect abnormal data batches over data streams with no fixed ending point. This paper is organized as follows to achieve these two aims. Section~\ref{sec:med_QIF} provides both algorithms and theoretical guarantees for our RenewQIF method. Section~\ref{sec:computing} discusses an extended Lambda architecture with an addition of quality control layer and pseudo code for numerical implementation, together with an analysis on algorithmic convergence and a monitoring procedure for abnormal data batches. Section~\ref{sec:sim_QIF} includes simulation results with comparisons of the proposed RenewQIF to the offline GEE, QIF and renewable GEE (RenewGEE) with or without abnormal data batches. Section~\ref{sec:realdata} illustrates the proposed method by a real data analysis application. The proofs of the large-sample properties for the RenewQIF method are included in the appendix. Derivation of RenewGEE as well as additional numerical results are included in the Supplementary Material.

\section{RenewQIF methodology}\label{sec:med_QIF}
\subsection{Offline QIF}\label{ssec:setup}
Consider independent streaming data batches consisting of cluster-correlated outcomes, sequentially generated from a common underlying population-averaged model~\citep{PAmodel1988} or marginal generalized linear model (MGLM)~\citep[Chapter 5]{Song2007correlated} with an unknown regression parameter $\hm{\beta}_0\in\Theta\subset\mathbb{R}^p$ where $\Theta$ is the parameter space for $\bm{\beta}$. For the ease of exposition, we assume an equal cluster size $m_i=m$. Our goal is to evaluate population-average effects of $p$ covariates, denoted by $\hm{\beta}_0=(\beta_{01},\dots,\beta_{0p})^\top$ in an MGLM with the marginal mean and covariance given by
\begin{equation}\label{eq:model_formula}
\hm{\mu}=\mathbb{E}(\hm{y} \mid \hm{X})=\left[ h(\hm{x}_{1}^\top\hm{\beta}_0),\dots,h(\hm{x}_{m}^\top\hm{\beta}_0) \right]^\top, \ \text{cov}(\hm{y} \mid \hm{X})=\phi \hm{\Sigma}(\hm{\beta}_0,\hm{\alpha})=\phi \hm{A}^{1/2}\hm{R}(\hm{\alpha})\hm{A}^{1/2},
\end{equation}
where
$\bm{y}=(y_1,\dots,y_m)^\top$, $\hm{\mu}=(\mu_1,\dots,\mu_m)^\top$ with $\mu_k=h(\hm{x}_k^\top\hm{\beta}_0)$ where $h(\cdot)$ is a known link function, and $\bm{X}=(\bm{x}_1,\dots,\bm{x}_m)^\top$ with $\hm{x}_k=(x_{k1},\dots,x_{kp})^\top$, $k=1,\dots,m$. $\phi$ is a dispersion parameter, $\hm{A} =\text{diag}\left\{ v(\mu_{1}),\dots,v(\mu_{m}) \right\}$ is a diagonal matrix with $v(\cdot)$ being a known variance function, and $\hm{R}(\hm{\alpha})$ is a working correlation matrix that is fully characterized by a correlation parameter vector $\hm{\alpha}$.

In the context of streaming data, consider a time point $b\geq 2$ with a total of $N_b$ clusters arriving sequentially in $b$ data batches, $\{\mathcal{D}_1,\dots,\mathcal{D}_b\}$, each containing $n_j=|\mathcal{D}_j|$, $j=1,\dots,b$, clusters. Let $\mathcal{D}_b^\star=\mathcal{D}_1\cup\dots\cup\mathcal{D}_b$ denote the cumulative collection of datasets up to data batch $b$ where each sample unit corresponds to an $m$-element vector of cluster-correlated outcomes, and the cumulative sample size is $N_b=|\mathcal{D}_b^\star|$. For simplicity, $\mathcal{D}_b$ (a single data batch $b$) or $\mathcal{D}_b^\star$ (an aggregation of $b$ data batches) is also used as respective sets of indices for clusters involved. For cluster $i$, let $\hm{y}_i=(y_{i1},\dots,y_{im})^\top$ and $\hm{X}_i=(\hm{x}_{i1},\dots,\hm{x}_{im})^\top$ be the correlated response vectors and associated covariates, $i=1,\dots,n_j$, $j=1,\dots,b$. According to~\citet{GEE1986}, an offline GEE estimator of $\hm{\beta}_0$ is a solution to the following generalized estimating equation for the cumulative data $\mathcal{D}_b^\star$ up to time point $b$:
\begin{equation}\label{eq:GEE_b}
\hm{\psi}_b^\star(\mathcal{D}_b^\star;\hm{\beta},\alpha)
=\sum_{i\in \mathcal{D}_b^\star}\hm{D}_{i}^\top \hm{\Sigma}_{i}^{-1}(\hm{y}_{i}-\hm{\mu}_{i})
=\hm{0},
\end{equation}
where $\hm{\mu}_{i}=(\mu_{i1},\dots,\mu_{im})^\top$, $\hm{D}_{i}=\partial \hm{\mu}_{i}/\partial \hm{\beta}^\top$ is an $m\times p$ matrix and $\hm{\Sigma}_i=\hm{A}_i^{1/2}\hm{R}(\hm{\alpha})\hm{A}_i^{1/2}$ with $\hm{A}_i=\text{diag}\left\{v(\mu_{i1}),\dots,v(\mu_{im}) \right\}$. According to~\citet{Qu2000}, the formulation of an offline QIF is based on an approximation to the inverse working correlation matrix by $\hm{R}^{-1}(\hm{\alpha})\approx \sum_{s=1}^{S}\gamma_{s}\hm{M}_{s}$, where $\gamma_{1},\dots,\gamma_{S}$ are constants possibly dependent on $\hm{\alpha}$, and $\hm{M}_{1},\dots,\hm{M}_{S}$ are known basis matrices with elements $0$ and $1$, which are determined by a given correlation matrix $\hm{R}(\hm{\alpha})$. In some cases, the above expansion can be exact. For example, as discussed in~\citet{Qu2000} and~\citet[Chapter 5]{Song2007correlated}, the basis matrices for the compound symmetry working correlation matrix are $\hm{M}_1=\hm{I}_{m}$, the identity matrix, and $\hm{M}_2^{cs}$, a matrix with all $0$ on the diagonal and all $1$ off the diagonal. %For AR-1 working correlation, three basis matrices include $\hm{M}_1=\hm{I}_{m}$, $\hm{M}_2^{ar}$ with $1$ on the two main off-diagonals and $0$ elsewhere, and $\hm{M}_3$ with $1$ on the corners $(1,1)$ and $(m,m)$, and $0$ elsewhere. 
Plugging such expansion into~\eqref{eq:GEE_b} leads to 
$
\hm{\psi}_b^\star(\mathcal{D}_b^\star;\hm{\beta},\alpha)=\sum_{i\in \mathcal{D}_b^\star}\sum_{s=1}^{S}\gamma_{s}\hm{D}_{i}^\top\hm{A}_{i}^{-1/2}\hm{M}_{s}\hm{A}_{i}^{-1/2}(\hm{y}_{i}-\hm{\mu}_{i})=\hm{0}$,
which may be regarded as a combination of the following extended score vector of $pS$ dimension:
\begin{equation*}
\begin{aligned}
\hm{g}_b^\star(\hm{\beta})=\sum_{i\in \mathcal{D}_b^\star}\hm{g}(\hm{y}_i;\hm{X}_i,\hm{\beta})
=\sum_{i\in \mathcal{D}_b^\star}
\begin{pmatrix}
\hm{D}_{i}^\top \hm{A}_{i}^{-1/2} \hm{M}_{1} \hm{A}_{i}^{-1/2}(\hm{y}_{i}-\hm{\mu}_{i}) \\
\vdots \\
\hm{D}_{i}^\top \hm{A}_{i}^{-1/2} \hm{M}_{S} \hm{A}_{i}^{-1/2}(\hm{y}_{i}-\hm{\mu}_{i})
\end{pmatrix}.
\end{aligned}
\end{equation*}
This is an over-identified estimating function, namely $dim(\hm{g}_b^\star(\hm{\beta})) > dim(\hm{\beta})$. To obtain an estimator of $\hm{\beta}_0$, following~\citet{Hansen1982}'s generalized method of moments (GMM), we take $\hm{\hat{\beta}}_b^\star=\underset{\hm{\beta}\in\mathbb{R}^p}{\arg \text{min}}\ Q_b^\star(\hm{\beta})$ with
\begin{equation}\label{eq:Q_stat_1}
Q_b^\star(\hm{\beta})= \hm{g}_b^{\star}(\hm{\beta})^\top \left\{\hm{C}_b^\star(\hm{\beta})\right\}^{-1} \hm{g}_b^\star(\hm{\beta}),
\end{equation}
where $\hm{C}_b^\star(\hm{\beta})=\sum_{i\in \mathcal{D}_b^\star}\hm{g}(\hm{y}_i;\hm{X}_i,\hm{\beta}) \hm{g}(\hm{y}_i;\hm{X}_i,\hm{\beta})^\top$ is the sample variance of $\hm{g}_b^\star(\hm{\beta})$. Note that the nuisance correlation parameter $\hm{\alpha}$ is not involved in~\eqref{eq:Q_stat_1} for the estimation of $\hm{\hat{\beta}}_b^\star$. 

\subsection{Online QIF}
Instead of processing the cumulative data $\mathcal{D}_b^\star$ once to obtain an offline QIF as shown above, we may conduct the online estimation and inference via a sequential and recursive updating scheme. In the proposed online estimation framework, let $\hm{\tilde{\beta}}_b$ be a renewable estimator, which is initialized by $\hm{\tilde{\beta}}_1=\hm{\hat{\beta}}_1=\underset{\hm{\beta}\in\mathbb{R}^p}{\arg\text{min}}\ Q_1(\hm{\beta})$, namely the QIF estimate obtained with the first data batch. When data batch $\mathcal{D}_b$ arrives,  a previous estimator $\hm{\tilde{\beta}}_{b-1}$ is renewed or updated to $\hm{\tilde{\beta}}_b$ using historical summary statistics of previous data batches $\mathcal{D}_{b-1}^\star$ and full observations of current data batch $\mathcal{D}_b$. This sequentially renewed estimator is termed as the renewable QIF or RenewQIF. After the completion of this updating, individual-level data of $\mathcal{D}_b$ is no longer accessible for the sake of data storage, but only the updated estimate $\hm{\tilde{\beta}}_{b}$ and summary statistics are carried forward in future calculations. For the empirical version, we use $\hm{g}_b(\hm{\beta}) = \sum_{i \in \mathcal{D}_b}\hm{g}(\hm{y}_i;\hm{X}_i,\hm{\beta})$ to denote the extended score vector of data batch $\mathcal{D}_b$, clearly $\hm{g}_b^\star(\hm{\beta}) = \sum_{j=1}^{b}\hm{g}_j(\hm{\beta})$. The negative gradient and sample variance matrix of $\hm{g}_b(\hm{\beta})$ are denoted by $\hm{G}_b(\hm{\beta})=-\sum_{i\in \mathcal{D}_b}\partial \hm{g}(\hm{y}_i;\hm{X}_i,\hm{\beta}) /\partial \hm{\beta}^\top$ and $\hm{C}_b(\hm{\beta})=\sum_{i\in \mathcal{D}_b}\hm{g}(\hm{y}_i;\hm{X}_i,\hm{\beta})\hm{g}(\hm{y}_i;\hm{X}_i,\hm{\beta})^\top$, respectively. In the theoretical framework, let the population variability matrix and sensitivity matrix be  $\mathbb{C}(\hm{\beta})=\mathbb{E}_{\hm{\beta}}\left\{\hm{g}(\hm{y};\hm{X},\hm{\beta})\hm{g}^\top(\hm{y};\hm{X},\hm{\beta})  \right\}$ and $\mathbb{G}(\hm{\beta}) = \mathbb{E}_{\hm{\beta}}\left\{ -\partial \hm{g}(\hm{y};\hm{X},\hm{\beta})/\partial\hm{\beta}^\top \right\}$.
In the process of RenewQIF, the same basis matrices are used in all data batches.

We begin the derivation of RenewQIF with two batches, the second one $\mathcal{D}_2$ arriving after the first $\mathcal{D}_1$. This simple scenario can be easily generalized to the case of an arbitrary number of data batches with little effort. According to~\citet{Qu2000}, a QIF estimator, $\hm{\hat{\beta}}_1=\underset{\hm{\beta}\in\mathbb{R}^p}{\arg \text{min}}\ Q_1(\hm{\beta})$ with $Q_1(\hm{\beta})=\hm{g}_1^\top(\hm{\beta}) \left\{\hm{C}_1(\hm{\beta}) \right\}^{-1} \hm{g}_1(\hm{\beta})$, may be obtained from the estimating equation:
$\hm{G}_1^\top(\hm{\hat{\beta}}_1) \left\{\hm{C}_1(\hm{\hat{\beta}}_1)\right\}^{-1}\hm{g}_1(\hm{\hat{\beta}}_1)=\hm{0}$. When $\mathcal{D}_2$ arrives, we aim to obtain the offline QIF estimator, $\hm{\hat{\beta}}_2^\star$, based on the accumulated data $\mathcal{D}_2^\star$, satisfying the estimating equation $\hm{G}_2^\star(\hm{\hat{\beta}}_2^\star)^\top\hm{C}_2^\star(\hm{\hat{\beta}}_2^\star)^{-1}\hm{g}_2^\star(\hm{\hat{\beta}}_2^\star)=\hm{0}$, or equivalently
\begin{equation}\label{eq:QIF_2}
\left\{\hm{G}_1(\hm{\hat{\beta}}_2^\star)+\hm{G}_2(\hm{\hat{\beta}}_2^\star) \right\}^\top
\left\{\hm{C}_1(\hm{\hat{\beta}}_2^\star)+\hm{C}_2(\hm{\hat{\beta}}_2^\star) \right\}^{-1}
\left\{\hm{g}_1(\hm{\hat{\beta}}_2^\star)+\hm{g}_2(\hm{\hat{\beta}}_2^\star) \right\}=\hm{0}.
\end{equation} 
Clearly, solving~\eqref{eq:QIF_2} for $\hm{\hat{\beta}}_2^\star$ involves subject-level data from both batches $\mathcal{D}_1$ and $\mathcal{D}_2$ where $\mathcal{D}_1$ may no longer be accessible. Our RenewQIF estimation is able to handle this issue. To proceed, heuristically, we may take the first-order Taylor expansions of the terms $\hm{g}_1(\hm{\hat{\beta}}_2^\star)$, $\hm{G}_1(\hm{\hat{\beta}}_2^\star)$ and $\hm{C}_1(\hm{\hat{\beta}}_2^\star)$ element-wise in~\eqref{eq:QIF_2} around $\hm{\hat{\beta}}_1$. Given that $\bm{g}_1$ is continuously differentiable and $\bm{G}_1$ is Lipschitz continuous in $\Theta$, it follows that
$
\frac{n_1}{N_2}\hm{g}_1(\hm{\hat{\beta}}_2^\star)
=\frac{n_1}{N_2}\hm{g}_1(\hm{\hat{\beta}}_1)+\frac{n_1}{N_2}\hm{G}_1(\hm{\hat{\beta}}_1)(\hm{\hat{\beta}}_1-\hm{\hat{\beta}}_2^\star)+O_p\left(\frac{n_1}{N_2}\|\hm{\hat{\beta}}_1-\hm{\hat{\beta}}_2^\star \|^2\right)$,
$\frac{n_1}{N_2}\hm{G}_1(\hm{\hat{\beta}}_2^\star)=\frac{n_1}{N_2}\hm{G}_1(\hm{\hat{\beta}}_1)+ O_p\left(\frac{n_1}{N_2}\|\hm{\hat{\beta}}_1-\hm{\hat{\beta}}_2^\star \|\right)$, and
$\frac{n_1}{N_2}\hm{C}_1(\hm{\hat{\beta}}_2^\star) =\frac{n_1}{N_2}\hm{C}_1(\hm{\hat{\beta}}_1)+O_p\left(\frac{n_1}{N_2}\|\hm{\hat{\beta}}_1-\hm{\hat{\beta}}_2^\star \|\right)$.
The error terms $O_p\left(\frac{n_1}{N_2}\|\hm{\hat{\beta}}_1-\hm{\hat{\beta}}_2^\star \|\right)$ and $O_p\left(\frac{n_1}{N_2}\|\hm{\hat{\beta}}_1-\hm{\hat{\beta}}_2^\star \|^2\right)$ may be asymptotically ignorable if $N_2$ is large enough. Dropping such error terms, we propose a new QIF estimator $\hm{\tilde{\beta}}_2$ as a solution to the estimating equation: $\hm{\tilde{G}}_2(\hm{\tilde{\beta}}_2)^\top \hm{\tilde{C}}_2(\hm{\tilde{\beta}}_2)^{-1}
\hm{\tilde{g}}_2(\hm{\tilde{\beta}}_2)=\hm{0}$, or equivalently, 
\begin{equation}\label{eq:renew_2}
\begin{aligned}
 \left\{\hm{G}_1(\hm{\hat{\beta}}_1)+\hm{G}_2(\hm{\tilde{\beta}}_2) \right\}^\top
\left\{\hm{C}_1(\hm{\hat{\beta}}_1)+\hm{C}_2(\hm{\tilde{\beta}}_2) \right\}^{-1}
\left\{\hm{g}_1(\hm{\hat{\beta}}_1)+\hm{G}_1(\hm{\hat{\beta}}_1)(\hm{\hat{\beta}}_1 - \hm{\tilde{\beta}}_2)+\hm{g}_2(\hm{\tilde{\beta}}_2) \right\} =\hm{0},
\end{aligned}
\end{equation}
where $\hm{\tilde{g}}_2$, $\hm{\tilde{G}}_2$ and $\hm{\tilde{C}}_2$ are, respectively, the resulting adjusted extended score vector, the online aggregated negative gradient, and the online sample variance matrix, none of which is calculated with the subject-level raw data $\mathcal{D}_1$. Thus, equation~\eqref{eq:renew_2} updates the initial $\hm{\hat{\beta}}_1$ to a new estimate with no use of the raw data in $\mathcal{D}_1$. Thus, $\hm{\tilde{\beta}}_2$ is called {\em a RenewQIF estimator} of $\hm{\beta}_0$, and equation~\eqref{eq:renew_2} is termed as {\em an incremental QIF estimating equation}. Furthermore, it is straightforward to find the RenewQIF estimator $\hm{\tilde{\beta}}_2$ numerically via the Newton-Raphson algorithm. That is, at the $(r+1)$-th iteration,
\begin{equation*}
%\begin{split}
\hm{\tilde{\beta}}_2^{(r+1)}=\hm{\tilde{\beta}}_2^{(r)}
+\left\{\hm{\tilde{G}}_2(\hm{\tilde{\beta}}_2^{(r)})^\top \hm{\tilde{C}}_2(\hm{\tilde{\beta}}_2^{(r)})^{-1} \hm{\tilde{G}}_2(\hm{\tilde{\beta}}_2^{(r)}) \right\}^{-1}
\hm{\tilde{G}}_2(\hm{\tilde{\beta}}_2^{(r)})^\top
\hm{\tilde{C}}_2(\hm{\tilde{\beta}}_2^{(r)})^{-1}
\hm{\tilde{g}}_2(\hm{\tilde{\beta}}_2^{(r)}),
%\end{split}
\end{equation*}
where $\hm{\tilde{G}}_2(\hm{\tilde{\beta}}_2^{(r)})=\hm{G}_1(\hm{\hat{\beta}}_1)+\hm{G}_2(\hm{\tilde{\beta}}_2^{(r)})$ and $\hm{\tilde{C}}_2(\hm{\tilde{\beta}}_2^{(r)})=\hm{C}_1(\hm{\hat{\beta}}_1)+\hm{C}_2(\hm{\tilde{\beta}}_2^{(r)})$. %{\color{blue}Here, the gradient is also an approximation that is traditionally used in the offline QIF with higher-order terms being dropped~\citep{Qu2000}.} 
Once again, it is worth pointing out that the above iterations do not require the subject-level data of $\mathcal{D}_1$, but only the historical summary statistics, including estimate $\hm{\hat{\beta}}_1$, its negative gradient $\hm{G}_1(\hm{\hat{\beta}}_1)$ and sample variance matrix $\hm{C}_1(\hm{\hat{\beta}}_1)$. In the QIF estimation above, the nuisance correlation parameter $\hm{\alpha}$ is not involved in the iterations, either.

Extending the above renewable procedure to a general setting of streaming datasets, we now define a renewable QIF estimation of $\hm{\beta}_0$ as follows. Let $\hm{\hat{\beta}}_b^\star$ be the offline QIF estimator of $\hm{\beta}_0$ with the cumulative data $\mathcal{D}_b^\star=\cup_{j=1}^b \mathcal{D}_j$ obtained from the offline QIF estimating equation $\hm{G}^\star_b(\hm{\hat{\beta}}_b^\star)^\top \hm{C}_b^\star(\hm{\hat{\beta}}_b^\star)^{-1} \hm{g}_b^\star(\hm{\hat{\beta}}_b^\star)=\hm{0}$. A renewable estimator $\hm{\tilde{\beta}}_b$ of $\hm{\beta}_0$ is defined as a solution to the incremental QIF estimating equation: $\hm{\tilde{G}}_b(\hm{\tilde{\beta}}_b)^\top \hm{\tilde{C}}_b(\hm{\tilde{\beta}}_b)^{-1} \hm{\tilde{g}}_b(\hm{\tilde{\beta}}_b)=\hm{0}$, which is equivalent to 
\begin{equation}\label{eq:renew_b_QIF}
\begin{aligned}
\bm{f}_b(\bm{\tilde{\beta}}_b) &= \left\{\sum_{j=1}^{b-1}\hm{G}_j(\hm{\tilde{\beta}}_j)+\hm{G}_b(\hm{\tilde{\beta}}_b)  \right\}^\top
\left\{\sum_{j=1}^{b-1}\hm{C}_j(\hm{\tilde{\beta}}_j)+\hm{C}_b(\hm{\tilde{\beta}}_b)  \right\}^{-1}\\
&\times \left\{\hm{\tilde{g}}_{b-1}(\hm{\tilde{\beta}}_{b-1})+
\sum_{j=1}^{b-1}\hm{G}_j(\hm{\tilde{\beta}}_j)(\hm{\tilde{\beta}}_{b-1}-\hm{\tilde{\beta}}_b)+\hm{g}_b(\hm{\tilde{\beta}}_b) \right\}=\hm{0},
\end{aligned}
\end{equation}
where $\hm{\tilde{G}}_b=\sum_{j=1}^{b}\hm{G}_j(\hm{\tilde{\beta}}_j)$ is the sequentially aggregated negative gradient matrix, $\hm{\tilde{C}}_b=\sum_{j=1}^{b}\hm{C}_j(\hm{\tilde{\beta}}_j)$ is the sequentially aggregated sample variance matrix, and $\bm{\tilde{g}}_b(\bm{\tilde{\beta}}_b)=\bm{\tilde{g}}_{b-1}(\bm{\tilde{\beta}}_{b-1})+\sum_{j=1}^{b-1}\hm{G}_j(\hm{\tilde{\beta}}_j)(\hm{\tilde{\beta}}_{b-1}-\hm{\tilde{\beta}}_b)+\hm{g}_b(\hm{\tilde{\beta}}_b)$ is the adjusted extended score vector. In effect, equation~\eqref{eq:renew_b_QIF} may be rewritten as a recursive formula:
\begin{equation}\label{eq:SGD_form}
\bm{\tilde{\beta}}_{b} = \bm{\tilde{\beta}}_{b-1} +N_b^{-1}\bm{{H}}_b^{-1}\bm{\tilde{U}}_b(\bm{\tilde{\beta}}_b),~\mbox{with}~\bm{\tilde{U}}_b(\bm{\tilde{\beta}}_b)=\bm{\tilde{G}}_b^\top\bm{\tilde{C}}_b^{-1}\left(\bm{\tilde{g}}_{b-1}+\bm{g}_b(\bm{\tilde{\beta}}_b)\right),
\end{equation}
where $\bm{{H}}_b=N_b^{-1}\bm{\tilde{G}}_b^\top\bm{\tilde{C}}_b^{-1}\bm{\tilde{G}}_{b-1}$. % and $\bm{\tilde{U}}_b(\bm{\tilde{\beta}}_b)=\bm{\tilde{G}}_b^\top\bm{\tilde{C}}_b^{-1}\left(\bm{\tilde{g}}_{b-1}+\bm{g}_b(\bm{\tilde{\beta}}_b)\right)$. 
Even though the first expression in~\eqref{eq:SGD_form} takes a similar form to that of the traditional SGD~\citep{Robbins1951,Sakrison1965,Toulis2017}, it is not an SGD as $\bm{\tilde{\beta}}_b$ is involved in both $\bm{\tilde{U}}_b(\cdot)$ and $\bm{H}_b$. It may be regarded as a second-order method involving Hessian inversion calculation useful for statistical inference. In comparison to SGD,~\eqref{eq:SGD_form} not only achieves efficient estimation but also provides relevant inferential quantities; the latter is of great importance in statistical analyses of cluster-correlated data streams. This RenewQIF estimate $\bm{\tilde{\beta}}_b$ given in~\eqref{eq:SGD_form} is clearly different from the offline QIF that requires to access the entire cumulative data $\mathcal{D}_b^\star$ to obtain the parameter estimate $\bm{\hat{\beta}}_b^\star$, and the latter may not be computationally feasible when the number of data batches $b\to\infty$.

Solving~\eqref{eq:renew_b_QIF} can be easily carried out via the following Newton-Raphson iterations:
\begin{equation}\label{eq:renew_algorithm}
\hm{\tilde{\beta}}_b^{(r+1)}=\hm{\tilde{\beta}}_b^{(r)}+
\left\{\hm{\tilde{G}}_b(\hm{\tilde{\beta}}_b^{(r)})^\top
\hm{\tilde{C}}_b(\hm{\tilde{\beta}}_b^{(r)})^{-1}
\hm{\tilde{G}}_b(\hm{\tilde{\beta}}_b^{(r)}) \right\}^{-1}
\hm{\tilde{G}}_b(\hm{\tilde{\beta}}_b^{(r)})^\top
\hm{\tilde{C}}_b(\hm{\tilde{\beta}}_b^{(r)})^{-1}
\hm{\tilde{g}}_b(\hm{\tilde{\beta}}_b^{(r)}).
\end{equation}
In algorithm~\eqref{eq:renew_algorithm}, clearly we only use the subject-level data of current data batch $\mathcal{D}_b$ and summary statistics $\{\hm{\tilde{\beta}}_{b-1},\hm{\tilde{g}}_{b-1},\hm{\tilde{G}}_{b-1},\hm{\tilde{C}}_{b-1} \}$ from historical data batches up to $b-1$ rather than subject-level raw data of $\mathcal{D}_{b-1}^\star$. Thus, our proposed RenewQIF is indeed an online estimation procedure.

\subsection{Large Sample Properties}\label{ssec:theorems_QIF}
In our discussion of large sample properties, the cumulative sample size $N_b\to\infty$ may arise from one of the following three scenarios: (S1) $n_j$ is finite for $j=1,\dots,b$ but the number of data batches $b\to\infty$; (S2) the initial data batch size $n_1\to\infty$ (aggregate the first few batches to create a large initial data batch prior to the renewable updating), and subsequent batch sizes $n_j$'s may be either finite or infinite for $j=2,\dots,b$, but the number of data batches $b$ is finite; and (S3) the initial data batch size $n_1\to\infty$ and other $n_j$'s are either finite or infinite for $j=2,\dots,b$, and the number of data batches $b\to\infty$.

Following the regularity conditions given in~\citet{Hansen1982}'s theory of GMM, we postulate the regularity conditions to establish some key large sample properties for RenewQIF.
\begin{condenum}
\item The true parameter $\bm{\beta}_0$ lies in the interior of parameter space $\Theta\subset \mathbb{R}^p$, and the space $\Theta$ is compact. ~\label{C1:compact}
\item The extended score vector $\hm{g}$ is unbiased such that $\mathbb{E}_{\hm{\beta}}\left\{\hm{g}(\hm{y};\hm{X},\hm{\beta})\right\}=\hm{0}$ if and only if $\hm{\beta}=\hm{\beta}_0$;~\label{C2:bound_g} %and $\sup_{\bm{\beta}\in \Theta}\left[\mathbb{E}_{\bm{\beta}} \|\bm{g}(\bm{y};\bm{X},\bm{\beta}) \|\right]<\infty$.  
\item The extended score vector $\hm{g}(\hm{y};\hm{X},\hm{\beta})$ is continuously differentiable with respect to parameter $\bm{\beta}$; its negative gradient $\hm{G}(\hm{y};\hm{X},\hm{\beta})=-\partial \hm{g}(\hm{y};\hm{X},\hm{\beta})/\partial \hm{\beta}^\top$ is Lipschitz continuous and the sensitivity matrix $\mathbb{G}(\bm{\beta}) = \mathbb{E}_{\bm{\beta}} \hm{G}(\hm{y};\hm{X},\hm{\beta})$ is of full column rank for $\hm{\beta}\in \Theta$.~\label{C3:bound_G}
\item $\mathbb{E}_{\bm{\beta}}\left[\|\bm{g}(\bm{y};\bm{X},\bm{\beta})\|^2\right]<\infty$ for all $\bm{\beta}\in \Theta$.~\label{C4:bound_C}
\item The variability matrix $\mathbb{C}(\hm{\beta}) =\mathbb{E}_{\bm{\beta}} \bm{g}(\bm{y};\bm{X},\bm{\beta})\bm{g}(\bm{y};\bm{X},\bm{\beta})^\top$ is positive-definite for $\bm{\beta}\in\Theta$.~\label{C5:positive_C}
\end{condenum}

\begin{remark}
	(C1)-(C5) are mild conditions required to establish asymptotic consistency for scenarios S2 and S3. The Lipschitz continuity of negative Hessian in (C3) implies that $\|\bm{G}(\bm{\beta}_1) - \bm{G}(\bm{\beta}_2)\|\leq M(\bm{X})\|\bm{\beta}_1-\bm{\beta}_2 \|$ for $\bm{\beta}_1,\bm{\beta}_2\in \Theta$, where $M(\bm{X})$ is a vector of covariates $\bm{X}$ which are all bounded under a fixed design considered in this paper. Condition (C4) indicates that the sample variance matrix $N_b^{-1}\bm{\tilde{C}}_b$ is finite for all $\bm{\beta}\in\Theta$. (C5) of positive-definite $\mathbb{C}(\bm{\beta})$ on the whole parameter space $\Theta$ is required specifically in scenario S1, while in scenarios S2 and S3, (C5) may be relaxed to hold in a neighborhood with radius $o(\sqrt{n_1})$. 
	
	Note that we do not include the convergence condition on a sequence of weighting matrices as in~\citet{Hansen1982} because the online QIF uses the sample variance matrix where $N_b^{-1}\bm{\tilde{C}}_b\overset{a.s.}{\to}\mathbb{C}$ holds under $N_b$ $i.i.d.$ samples and conditions (C1)-(C4). 
\end{remark}

We first establish estimation consistency for the RenewQIF in scenarios (S2)-(S3). 
\begin{theorem}\label{thm:consist_QIF_23}
	In scenarios (S2)-(S3), under regularity conditions (C1)-(C5), the renewable estimator $\hm{\tilde{\beta}}_b$ given in~\eqref{eq:renew_b_QIF} is consistent, namely
$
	\bm{\tilde{\beta}}_b \overset{p}{\to} \bm{\beta}_0, \ \text{as} \ N_b=\sum_{j=1}^{b}n_j\to \infty.
$
\end{theorem}

\begin{theorem}\label{thm:normal_QIF23}
	In scenarios (S2)-(S3), under regularity conditions (C1)-(C5), the renewable estimator $\hm{\tilde{\beta}}_b$ in~\eqref{eq:renew_b_QIF} is asymptotically normally distributed, namely
	$
	\sqrt{N_b}(\hm{\tilde{\beta}}_b-\hm{\beta}_0) \overset{d}{\to} \mathcal{N}\left( \hm{0},{\mathbb{J}}^{-1}(\hm{\beta}_0)\right)$, as $N_b\to \infty,
$
	where Godambe information $\mathbb{J}(\hm{\beta}_0)=\mathbb{G}^\top(\hm{\beta}_0)\mathbb{C}^{-1}(\hm{\beta}_0)\mathbb{G}(\hm{\beta}_0)$. 
\end{theorem}
The proof of Theorem \ref{thm:consist_QIF_23} and~\ref{thm:normal_QIF23} are given in Appendix~\ref{app:pf_QIF2}, respectively. %Under the same regularity conditions, we can further establish the asymptotic normality for the RenewQIF in the two scenarios (S2)-(S3).  The proof of Theorem~\ref{thm:normal_QIF23} is given in Appendix~\ref{app:pf_QIF2}. 
To establish consistency and asymptotic normality for scenario (S1), we need the following additional conditions. 
\begin{condenum}
\item[(C6)] The matrix $\bm{G}_1(\bm{\beta})^\top\bm{C}_1(\bm{\beta})^{-1}\bm{G}_1(\bm{\beta})$ is positive-definite for $\bm{\beta}\in\Theta$.~\label{C6:D1_pd}
\item[(C7)] $\bm{H}_b=N_b^{-1}\bm{\tilde{G}}_b^\top\bm{\tilde{C}}_b^{-1}\bm{\tilde{G}}_{b-1}$ is positive-definite for $b\geq2$;~\label{C7:Hb_pd}
\item[(C8)] Sample size in each data batch is uniformly bounded, i.e. $\sup_b n_b <C$ for a positive constant $C<\infty$, and $b=1,2,\dots$.~\label{C8:finite_nb}
\end{condenum}
\begin{remark}
Positive-definiteness of $\bm{G}_1(\bm{\beta})^\top\bm{C}_1(\bm{\beta})^{-1}\bm{G}_1(\bm{\beta})$ is not a strong condition as long as $n_1$ is moderately large. $\bm{H}_b$ is the multiplication of the aggregated summary statistics, and this condition should be satisfied as $b$ increases. Condition (C8) is the condition on data batch size in scenario (S1) where data batch size is finite and $b\to\infty$.
\end{remark}

\begin{theorem}\label{thm:consist_QIF_1}
	In scenario (S1), under regularity conditions (C1)-(C7), the renewable estimator $\hm{\tilde{\beta}}_b$ given in~\eqref{eq:renew_b_QIF} is consistent, namely
$
	\bm{\tilde{\beta}}_b \overset{p}{\to} \bm{\beta}_0, \ \text{as} \ N_b=\sum_{j=1}^{b}n_j\to \infty.
$
\end{theorem}

\begin{theorem}\label{thm:normal_QIF1}
	In scenario (S1), under regularity conditions (C1)-(C8), the renewable estimator $\hm{\tilde{\beta}}_b$ given in~\eqref{eq:renew_b_QIF} is asymptotically normally distributed, namely,
	$
	\sqrt{N_b}(\hm{\tilde{\beta}}_b-\hm{\beta}_0) \overset{d}{\to} \mathcal{N}\left( \hm{0},{\mathbb{J}}^{-1}(\hm{\beta}_0)\right)$, as $N_b\to \infty,
$
	where Godambe information $\mathbb{J}(\hm{\beta}_0)=\mathbb{G}^\top(\hm{\beta}_0)\mathbb{C}^{-1}(\hm{\beta}_0)\mathbb{G}(\hm{\beta}_0)$. 
\end{theorem}
The proof of Theorems \ref{thm:normal_QIF1} and~\ref{thm:normal_QIF1} are given in Appendix~\ref{app:pf_QIF1}.

It is interesting to notice that the asymptotic covariance matrix of the renewable estimator $\hm{\tilde{\beta}}_b$ given in both Theorems~\ref{thm:normal_QIF1} and~\ref{thm:normal_QIF23} is exactly the same as that of the offline estimator $\hm{\hat{\beta}}_b^\star$. This implies that the proposed RenewQIF estimator achieves the same asymptotic distribution as the offline QIF estimator. With no access to any historical subject-level data in the computation, using only the online aggregated matrices $\hm{\tilde{G}}_b=\sum_{j=1}^{b}\hm{G}_j(\mathcal{D}_j;\hm{\tilde{\beta}}_j)$ and $\hm{\tilde{C}}_b=\sum_{j=1}^{b}\hm{C}_j(\mathcal{D}_j;\hm{\tilde{\beta}}_j)$, we can calculate the estimated asymptotic covariance matrix as $\widetilde{\hm{\Sigma}_b}(\hm{\beta}_0) = \left\{N_b^{-1}\hm{\tilde{J}}_b \right\}^{-1}=N_b \left\{\hm{\tilde{G}}_b^\top\hm{\tilde{C}}_b^{-1}\hm{\tilde{G}}_b  \right\}^{-1}$.
It follows that the online estimated asymptotic variance matrix for the RenewQIF $\hm{\tilde{\beta}}_b$ is
\begin{equation}\label{eq:est_var}
\hm{\tilde{V}}(\hm{\tilde{\beta}}_b) \coloneqq
\widetilde{\text{Var}}(\hm{\tilde{\beta}}_b)
=\frac{1}{N_b} \widetilde{\hm{\Sigma}_b}(\hm{\beta}_0)
=\left\{\hm{\tilde{G}}_{b}^\top \hm{\tilde{C}}_{b}^{-1}
\hm{\tilde{G}}_{b}\right\}^{-1}.
\end{equation}

	Positive-definiteness on $\mathbb{C}(\bm{\beta})$ is required by (C5) for all $\bm{\beta}\in\Theta$, but the sample version $N_b^{-1}\bm{\tilde{C}}_b$ may be non-invertible. In the implementation, we invoke a generalized inverse (Moore-Penrose inverse). Other possible remedy approaches include (i) a linear shrinkage estimator to replace the sample covariance matrix~\citep{Han-Song-2011}; (ii) a selection of moment conditions~\citep{Cho-Qu-2015}; and (iii) a principle component dimension reduction~\citep{Pearson-1901} to ensure that the components in $\bm{g}$ are not (nearly) linearly dependent.

In addition, it is noteworthy that in (S2) or (S3), our method can be used as an alternative to parallelized computing methods. However, with our method, the convergence rate is $O_p(N_b^{-1/2})$ which is based on the cumulative sample size $N_b$. This indicates a faster convergence rate than parallelized distributed estimation where the convergence rate is based on the sample size of the smallest single dataset~\citep{LingSong2017} $\underset{j}{\min} \ \sqrt{n_j}$.

\section{Computing and Monitoring}\label{sec:computing}
\subsection{Computing Implementation of RenewQIF}\label{ssec:imple_QIF}
\begin{figure}[ht]
	\centering
	\includegraphics[width=\linewidth]{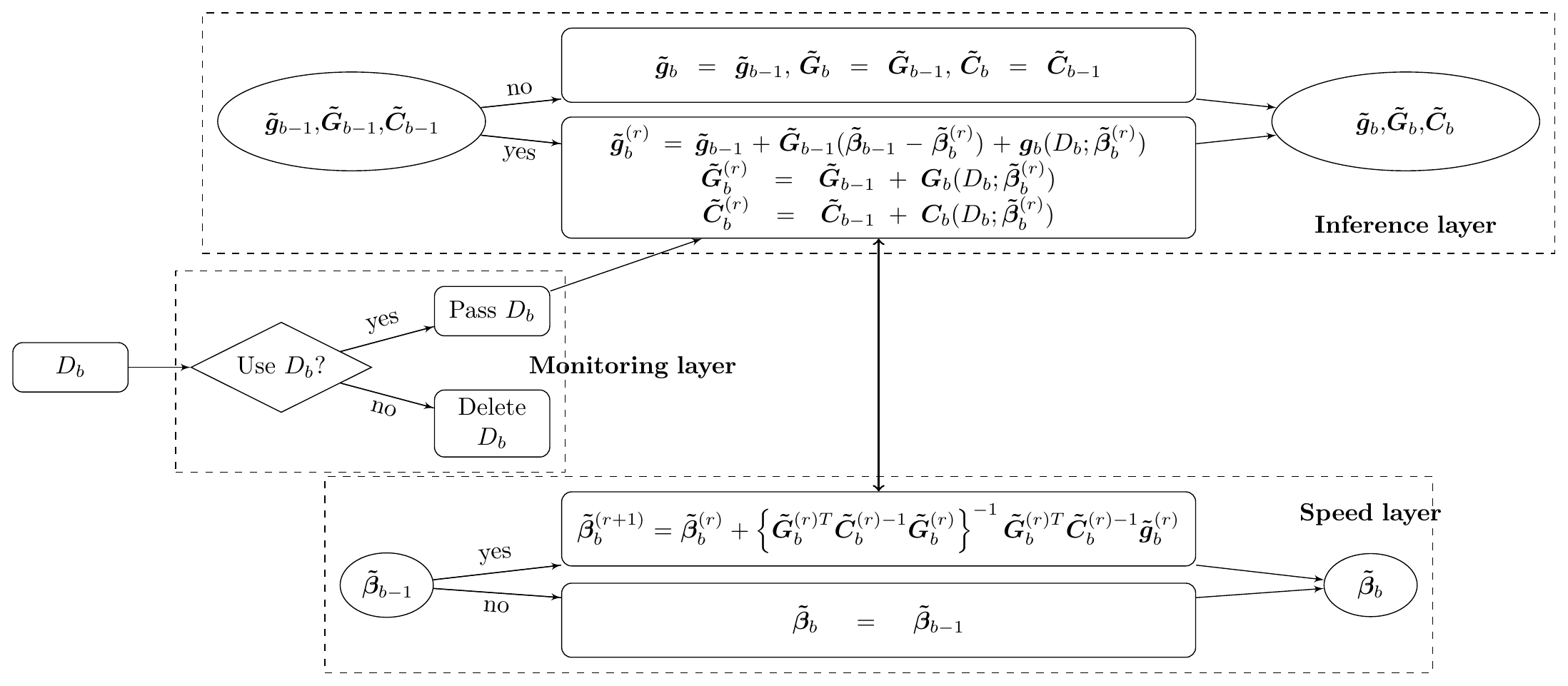}
	\vspace{-0.5cm}
	\caption{Diagram of an extended Lambda architecture with the addition of both monitoring and inference layers to the standard speed layer.}
	\label{fig:algorithmQIF}
\end{figure}

We expand the existing Spark's Lambda architecture to reduce computing burden in the proposed framework of RenewQIF methodology. The iterative calculation in~\eqref{eq:renew_algorithm} can be implemented in the speed and inference layers in an extended Lambda architecture shown in Figure~\ref{fig:algorithmQIF}. Here, relevant inferential statistics include the aggregated extended score vector $\hm{\tilde{g}}$ and two inferential matrices $\hm{\tilde{G}}$ (aggregated negative gradient) and $\hm{\tilde{C}}$ (aggregated sample variance matrix). If data batch $D_b$ passes the scrutiny, we update $\hm{\tilde{\beta}}_{b-1}$ to $\hm{\tilde{\beta}}_b$ at the speed layer and update $\hm{\tilde{g}}_{b-1}$, $\hm{\tilde{G}}_{b-1}$, $\hm{\tilde{C}}_{b-1}$ to $\hm{\tilde{g}}_{b}$, $\hm{\tilde{G}}_b$ and $\hm{\tilde{C}}_b$ at the inference layer. Otherwise, skip all updating steps and proceed to next data batch $\mathcal{D}_{b+1}$.  Algorithm~\ref{algorithm:code_QIF} lists the pseudo code for the implementation of the RenewQIF via the paradigm of the extended Lambda architecture shown in Figure~\ref{fig:algorithmQIF}. Some explanations are given below.

\clearpage
\begin{algorithm}[H]
	\caption{\label{algorithm:code_QIF} \small RenewQIF for streaming cluster-correlated data in the extended Lambda architecture.}
	\small
	%	\begin{algorithmic}
	\textbf{Inputs:} 
	\text{Sequentially arrived datasets} $\mathcal{D}_1$,...,$\mathcal{D}_b$,... from an {MGLM} with mean $\mathbb{E}(\bm{y}\mid\bm{X})=\hm{\mu}$ and covariance $\text{cov}(\bm{y}\mid\bm{X})=\phi\hm{\Sigma}$ specified in equation~\eqref{eq:model_formula}; \\
	\textbf{Outputs:}
	$\hm{\tilde{\beta}}_b$
	\text{and} $\hm{\tilde{V}}(\hm{\tilde{\beta}}_b)$, for $b=1,2,\dots$\ ; \\
	\textbf{Initialize:} Initial values $\hm{\tilde{\beta}}_{0}$, $\hm{\tilde{g}}_0=\hm{0}_{pS}$, $\hm{\tilde{G}}_0=\hm{0}_{pS\times p}$ and $\hm{\tilde{C}}_0=\hm{0}_{pS\times pS}$\ ; \\
	\For {$b=1,2,\dots$}{
		\text{Read in dataset} $\mathcal{D}_b$;  \\
		\text{At the monitoring layer, if $b\geq 2$, calculate}
		$\Lambda_b=Q_{1}(\hm{\check{\beta}}_b)+Q_b(\hm{\check{\beta}}_b)$; \\
		\text{if} $\Lambda_b\geq \chi^2_{df,\alpha}$, say $\alpha=0.05$, set $\hm{\tilde{\beta}}_b=\hm{\tilde{\beta}}_{b-1}$, $\hm{\tilde{g}}_b=\hm{\tilde{g}}_{b-1}$, $\hm{\tilde{G}}_b=\hm{\tilde{G}}_{b-1}$, $\hm{\tilde{C}}_b=\hm{\tilde{C}}_{b-1}$ \text{and jump to Line 16};\\
		{otherwise, start iterations with $\hm{\tilde{\beta}}_b$ initialized by $\hm{\tilde{\beta}}_{b-1}$;}\\ 
		\Repeat{converegence}{
			\text{At the inference layer, calculate} $\hm{\tilde{g}}_b^{(r)}=\hm{\tilde{g}}_{b-1}+\hm{\tilde{G}}_{b-1}(\hm{\tilde{\beta}}_{b-1}-\hm{\tilde{\beta}}_b)+\hm{g}_b(\mathcal{D}_b;\hm{\tilde{\beta}}_b^{(r)})$,\\
			$\hm{\tilde{G}}_b^{(r)}=\hm{\tilde{G}}_{b-1}+\hm{G}_b(\mathcal{D}_b;\hm{\tilde{\beta}}_b^{(r)})$ and
			$\hm{\tilde{C}}_b^{(r)}=\hm{\tilde{C}}_{b-1}+\hm{C}_b(\mathcal{D}_b;\hm{\tilde{\beta}}_b^{(r)})$ ; \\
			\text{At the speed layer,} $\hm{\tilde{\beta}}_b^{(r+1)}=\hm{\tilde{\beta}}_b^{(r)}+\left\{ \hm{\tilde{G}}_b^{(r)^\top}\hm{\tilde{C}}_b^{(r)^{-1}}\hm{\tilde{G}}_b^{(r)} \right\}^{-1} 
			\hm{\tilde{G}}_b^{(r)^\top}\hm{\tilde{C}}_b^{(r)^{-1}} \hm{\tilde{g}}_b^{(r)}$\ ;
		}
		\text{At the inference layer, calculate} $\hm{\tilde{V}}(\hm{\tilde{\beta}}_b)=\left\{ \hm{\tilde{G}}_b^\top\hm{\tilde{C}}_b^{-1}\hm{\tilde{G}}_b\right\}^{-1}$; \\
		\text{Save $\hm{\tilde{\beta}}_b$, $\hm{\tilde{g}}_b$, $\hm{\tilde{G}}_b$} and $\hm{\tilde{C}}_b$ at the speed and inference layers, respectively; \\
		Release dataset $\mathcal{D}_b$ from the memory.
	}
	\textbf{Return} $\hm{\tilde{\beta}}_b$ and $\hm{\tilde{V}}(\hm{\tilde{\beta}}_b)$, for $b=1,2,\dots$. 
\end{algorithm}

%\begin{enumerate}
	\noindent1. Line 1: the population-averaged model or MGLM has been specified in Section~\ref{ssec:setup}. \\
	2. Line 2: the outputs include RenewQIF estimates of the regression coefficients and the corresponding estimated asymptotic covariance matrix at each time point $b$, and the latter is needed for statistical inference.\\
	3. Line 3: set certain initial values for the regression coefficients, {\em e.g.}, get the initial estimate $\hm{\tilde{\beta}}_{0}$ by fitting $\mathcal{D}_1$ to R function~\texttt{glm()}. \\
	4. Line 4: run through the sequential updating procedure along data streams.\\
	5. Line 6: before updating $\bm{\tilde{\beta}}_{b-1}$ with current $\mathcal{D}_b$, first check its compatibility with the normal reference data batch $\mathcal{D}_{1}$. QIF estimator $\hm{\check{\beta}}_b$ is obtained by minimizing the quadratic inference function based only on these two data batches, $\mathcal{D}_{1}\cup \mathcal{D}_b$. The goodness-of-fit test statistic $\Lambda_b$ will be discussed in detail in Section~\ref{ssec:monitor}. \\
    6. Line 7: if the the current data batch $\mathcal{D}_b$ does not pass the scrutiny test, we don't use it in the updating and jump to Line 16 directly. \\
	7. Lines 8-11: if the test concludes the current data batch $\mathcal{D}_b$ passes the scrutiny, at the inference layer, utilize the prior estimate $\hm{\tilde{\beta}}_{b-1}$ and current batch $\mathcal{D}_b$ to calculate $\hm{\tilde{g}}_b$, $\hm{\tilde{G}}_b$ and $\hm{\tilde{C}}_b$ through a communication with the speed layer. \\
	8. Line 12: run the Newton-Raphson algorithm to renew $\hm{\tilde{\beta}}_{b-1}$ to $\hm{\tilde{\beta}}_b$. \\
    9. Line 13: the convergence criterion and the distance between theoretical estimator $\bm{\tilde{\beta}}_b$ and experimental estimator $\bm{\tilde{\beta}}_b^{(r)}$ will be further discussed in Section~\ref{ssec:converge}.\\ 
	10. Line 14: the inference layer performs statistical inference with $\hm{\tilde{G}}_b$ and $\hm{\tilde{C}}_b$. 
%\end{enumerate}

\begin{figure}[ht]
    \centering
	\includegraphics[width=0.9\linewidth]{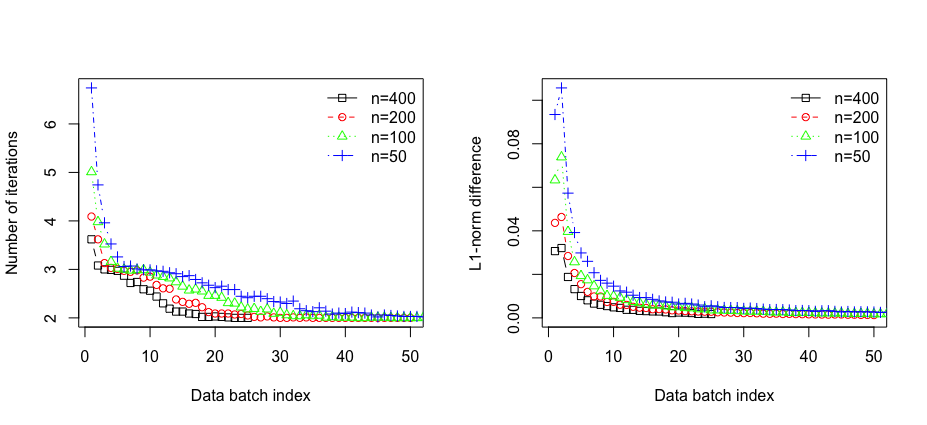}
	\vspace{-0.5cm}
	\caption{
	The left panel shows the number of iterations to reach $\Delta_r<10^{-6}$ with different data batch size $n_b$; the right panel indicates the $L_1$-norm difference $\|\bm{\tilde{\beta}}_b-\bm{\tilde{\beta}}_{b-1}\|$ decreases fast as $b$ increases. 
	Both plots are generated under a marginal logistic model specified in Section~\ref{ssec:setup_QIF} with a fixed $N_B=10^4$ but varying batch size $n_b$.}
	\label{fig:conv_analysis}
\end{figure}

\begin{figure}[h]
	\centering
		\includegraphics[width=\linewidth]{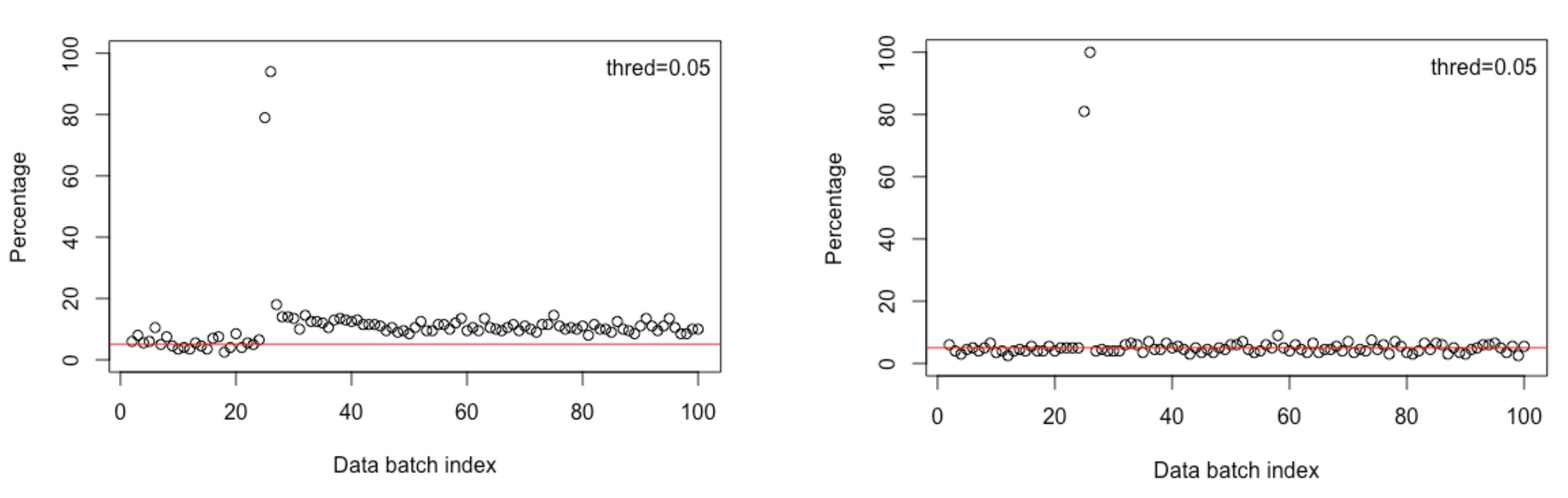}
	\caption{These two plots are generated under the marginal logistic model specified in Section~\ref{ssec:setup_QIF} with a fixed total sample size $N_B=10^4$ and data batch size $n_b=100$. Two abnormal data batches are $\mathcal{D}_{25}$ and $\mathcal{D}_{26}$. The $y$-axis is the empirical proportion of rejections over 500 replications, and $x$-axis is the index of test statistic $\Lambda_b$ (also the data batch index).}
	\label{fig:test}
\end{figure}

\subsection{Monitoring of Algorithmic Convergence}\label{ssec:converge}
Note that the large sample properties developed in Section~\ref{ssec:theorems_QIF} centers at the theoretical root, $\bm{\tilde{\beta}}_b$, our proposed incremental estimating equation in~\eqref{eq:renew_b_QIF}, i.e. $\bm{f}_b(\bm{\tilde{\beta}}_b)=\bm{0}$. In the implementation, $\bm{\tilde{\beta}}_b^{(r)}$ is the actual estimate harvested numerically at the $r$-th iteration of the Newton-Raphson algorithm~\eqref{eq:renew_algorithm}. Following the conventional practice, we set up a stringent convergence criterion, so that the two quantities $\bm{\tilde{\beta}}_b$ and $\bm{\tilde{\beta}}_b^{(r)}$ are numerically close enough that their difference is ignored. According to~\citet{Nocedal2006}[Chapter 3], the convergence rate of the algorithm~\eqref{eq:renew_algorithm} is at least quadratic if the following conditions are satisfied:
(a) the starting point $\bm{\tilde{\beta}}_b^{(0)}$ is sufficiently close to the root $\bm{\tilde{\beta}}_b$; (b) the negative gradient $\bm{\tilde{J}}_b=-\nabla \bm{f}_b(\bm{\beta}) = \bm{\tilde{G}}_b^\top\bm{\tilde{C}}_b^{-1}\bm{\tilde{G}}_b$ is Lipschitz continuous in a neighborhood of the root $\bm{\tilde{\beta}}_b$; and (c) $\bm{f}_b(\bm{\beta})$ is differentiable and  $\bm{\tilde{J}}_b$ is positive-definite in a neighborhood of the root $\bm{\tilde{\beta}}_b$. These three conditions are all satisfied by our algorithm~\eqref{eq:renew_algorithm}. This is because condition (a) holds for large $b$ in renewable updating due to the initialization of $\bm{\tilde{\beta}}_b^{(0)}=\bm{\tilde{\beta}}_{b-1}$. Being an example, the right panel of Figure~\ref{fig:conv_analysis} shows the closeness to the root $\bm{\tilde{\beta}}_b^{(r)}$ improves over the course of increased $b$. Conditions (b) and (c) are also satisfied by the regularity conditions~\ref{C3:bound_G} and~\ref{C5:positive_C} in Section~\ref{ssec:theorems_QIF}.

Line 13 in Algorithm~\ref{algorithm:code_QIF} stems from the convergence criteria in the Newton's decrement given by $\Delta_r = \bm{f}_b^\top(\bm{\tilde{\beta}}_b^{(r)}) \left(\bm{\tilde{J}}_b(\bm{\tilde{\beta}}_b^{(r)})\right)^{-1} \bm{f}_b(\bm{\tilde{\beta}}_b^{(r)}) < 10^{-6}$, which measures the proximity of $\bm{\tilde{\beta}}_b^{(r)}$ to $\bm{\tilde{\beta}}_b$. Clearly, $\Delta_r = 0$ when $\bm{\tilde{\beta}}^{(r)} = \bm{\tilde{\beta}}_b$ should be monitored over iterations of algorithm~\eqref{eq:renew_algorithm}. The algorithm will stop once $\Delta_r$ satisfies the stopping rule $\Delta_r< 10^{-6}$. To control the computation time, we also terminate the algorithm when the number of iterations reaches a pre-fixed threshold. Based on our extensive empirical experience, in our current implementation, we set the maximum number of iterations at $50$. If this threshold is reached with failure of $\Delta_r<10^{-6}$, a warning message ``algorithm reached `maxit' but did not reach the convergence criteria" will be given as an output. All simulation studies in Section~\ref{sec:sim_QIF} have shown this criterion is satisfactory with zero warning message. As shown in the left panel in Figure~\ref{fig:conv_analysis}, the number of iterations required to reach the criteria $\Delta_r<10^{-6}$ is all less than $10$, and with the sequential addition of data batches, the number of iterations run by algorithm~\eqref{eq:renew_algorithm} decreases monotonically as $b$ increases.

\subsection{Monitoring of Abnormal Data Batches}\label{ssec:monitor}
For the case of high throughput data streams in practice, it is highly likely to encounter abnormal data batches. To address this issue, we relax the RenewQIF method to a situation where abnormal data batches may occur over the sequence of data streams, $\mathcal{D}_2,\dots,\mathcal{D}_b$. A data batch $\mathcal{D}_\tau$, $\tau\in \{2,\dots,b\}$, is regarded as being abnormal if $\mathcal{D}_\tau$ is generated from a model whose regression parameters, say $\hm{\beta}_\tau$'s, are different from those of the underlying main model of interest, $\hm{\beta}_0$ of the true model, i.e. $\hm{\beta}_\tau\neq\hm{\beta}_0$. In other words, $\mathcal{D}_\tau$ is an outlying data batch, which is incompatible with the data batches generated from the true model. Let $\Gamma_q=\{\tau_1,\dots,\tau_q \}$ denote the set of indices for $q$ abnormal data batches. In reality, we do not know set $\Gamma_q$ in advance but want to find them out during the collection of data streams. For convenience, we assume that the first data batch $\mathcal{D}_1$ is the normal reference, which is generated from a model with ``the normal" parameter $\hm{\beta}_0$. At each subsequent time point $b$ ($b\geq 2$), we propose a diagnostic procedure via hypothesis testing of mean-zero assumption for the pair of extended scores $H_0: \mathbb{E}_{\hm{\beta}}(\hm{g}_{1})=\mathbb{E}_{\hm{\beta}}(\hm{g}_b)=\hm{0}$, where $\bm{g}_1$ and $\bm{g}_b$ are the extended score vectors for data batches $\mathcal{D}_1$ and $\mathcal{D}_b$, respectively, similar to the one given in~\eqref{eq:QIF_2}. This goodness-of-fit test essentially enables to check the compatibility between data batch $\mathcal{D}_b$ under investigation and the normal reference $\mathcal{D}_{1}$. If $H_0$ is rejected, we would not use data batch $\mathcal{D}_b$ to renew $\hm{\tilde{\beta}}_{b-1}$, and set $\hm{\tilde{\beta}}_{b}$ equal to $\hm{\tilde{\beta}}_{b-1}$; otherwise, execute an update from $\hm{\tilde{\beta}}_{b-1}$ to $\hm{\tilde{\beta}}_b$ using RenewQIF. Next we proceed to test $H_{0}$ with next data batch $\mathcal{D}_{b+1}$.

We construct a test statistic along the line of~\citet{Hansen1982}'s seminal goodness-of-fit test. The quadratic inference function has useful chi-squared properties for hypothesis testing~\citep{Qu2003}. For checking for data compatibility, we consider a quadratic inference function of the following form:
\begin{equation}\label{eq:test}
\begin{split}
\Lambda_b(\hm{\beta}) &= 
\begin{pmatrix}
\hm{g}_{1}(\hm{\beta}) \\
\hm{g}_{b}(\hm{\beta})
\end{pmatrix}^\top
\begin{pmatrix}
\hm{C}_{1}(\hm{\beta})  &\hm{0} \\
\hm{0}              &\hm{C}_b(\hm{\beta})
\end{pmatrix}^{-1}
\begin{pmatrix}
\hm{g}_{1}(\hm{\beta}) \\
\hm{g}_b(\hm{\beta})
\end{pmatrix}, 
\end{split}
\end{equation}
where $\hm{C}_{1}$ and $\hm{C}_b$ are the estimated sample covariances of extended scores $\hm{g}_{1}$ and $\hm{g}_b$, respectively. Note that the form of block-diagonal covariance in $\Lambda_b$ is due to the independence between $\mathcal{D}_{1}$ and $\mathcal{D}_b$. Let $\hm{\check{\beta}}_b = \underset{\hm{\beta} \in \mathbb{R}^p}{\arg \min} \ \Lambda_b(\hm{\beta})$, under $H_0$, i.e. $b \not\in \Gamma_q$, test statistic $\Lambda_b(\hm{\check{\beta}}_b) \xrightarrow[]{d} \chi_{df}^2$ with $df=rank(\hm{C}_{1})+rank(\hm{C}_{b})-p$; under $H_1$, for an index $\tau\in\Gamma_{q}$, and given local alternatives in the form of $\hm{\beta}_{\tau}=\hm{\beta}_0+(n_{1}+n_\tau)^{-1/2}\hm{d}$, $\hm{d}\in\mathbb{R}^p$, test statistic $\Lambda_\tau(\hm{\check{\beta}}_\tau)
\xrightarrow[]{d} \chi_{df}^2(\lambda)$, with $df=rank(\hm{C}_{1})+rank(\hm{C}_{\tau})-p$ and the non-centrality parameter $\lambda= \hm{d}^\top \mathbb{J}(\hm{\beta}_0) \hm{d}$ where $\mathbb{J}$ is the Godambe information matrix given in Theorems~\ref{thm:normal_QIF1} and~\ref{thm:normal_QIF23}. Moreover, it is easy to show that $\text{Power} = P_{H_1}\left(
\Lambda_\tau(\hm{\check{\beta}}_{\tau}) >\chi_{df,\alpha}^2\right)\to1$, as $(n_{1}+n_\tau)\to \infty$, which implies that the proposed test $\Lambda_\tau$ is consistent. Under a finite sample size, with fixed $\hm{d}$, the power of $\Lambda_\tau$ depends on both statistical significance level $\alpha$ and abnormal data batch size $n_\tau$. Larger $\alpha$ leads to higher power and smaller type II error, but also a higher chance to produce false alarms. Obviously, increasing data batch size $n_b$ will help increase power. The above monitoring procedure is based on the asymptotic properties under scenarios S2 and S3 with $n_j\to\infty$ for some $j$. However, if the reference data batch size is small as assumed in scenario S1, one may carry out the proposed diagnostic test against an augmented reference data  that combines several normal data batches to reach a desirable sample size.

In practice the use of  the first data batch $\mathcal{D}_1$, which is the normal reference sampled from the true model, may be replaced by any data batch that is deemed appropriate.  This choice is obviously somewhat subjective and mostly made by practitioners base on their prior experience and existing knowledge on data quality.  We do not recommend frequently changing the reference batch, but combining several normal data batches to form a larger reference one is useful to reach more stable performance of the diagnostic test $\Lambda_b$.  For example, adaptively using the adjacent data batch as the reference,  we show in the left panel of Figure~\ref{fig:test} that the monitoring diagnostic test suffers from an inflated type I error once an abnormal data batch is mistakenly set as the reference. Therefore, fixing a single or an augmented reference data batch in the monitoring leads to reliable diagnoses for abnormal data batches, as shown in the right panel of Figure~\ref{fig:test} with the normal reference fixed at the first data batch.

\section{Simulation Experiments}\label{sec:sim_QIF}
\subsection{Setup}\label{ssec:setup_QIF}
We conduct simulation experiments to assess the performances of the proposed RenewQIF estimation and inference, as well as of the diagnostic procedure for abnormal data batches, in the setting of marginal generalized linear models (MGLMs) for cluster-correlated data streams. We compare the proposed RenewQIF method with (i) the offline GEE estimator obtained by processing the entire cumulative data once, (ii) the offline QIF estimator obtained by processing the entire data once, and (iii) renewable GEE estimation method (RenewGEE) that is similar to RenewQIF (see the relevant derivation in Section 2 in the Supplementary Material).

In the first part of comparisons to be presented below, we consider the following criterion related to both parameter estimation and inference: (a) averaged absolute bias (A.bias), (b) averaged asymptotic standard error (ASE), (c) empirical standard error (ESE), and (d) coverage probability (CP). Both offline GEE and offline QIF estimates are yielded from the R packages \texttt{gee} and \texttt{qif}. Computational efficiency is assessed by (e) computation time (C.Time) and (f) running time (R.Time). R.Time accounts only algorithm execution time, while C.Time includes time spent on both data loading and algorithm execution. In the second part of comparisons, we will first evaluate the type I error and power of the proposed goodness-of-fit test for data compatibility with different significance level $\alpha$ and data batch size $n_b$. In addition, the criteria for parameter estimation and inference will be reported thoroughly on the competing methods with and without quality control. 

In the simulation studies, we set a terminal point $B$, and generate a full dataset $\mathcal{D}_B^\star$ with $N_B$ independent cluster-correlated observations of an $m$-dimensional MGLM, consisting of marginal mean $\mathbb{E}(\hm{y}_{i}\mid\hm{X}_{i})=\left[h(\hm{x}_{i1}^\top\hm{\beta}_0),\dots,h(\hm{x}_{im}^\top\hm{\beta}_0) \right]^\top$ with $\hm{\beta}_0=(0.2,-0.2,0.2,-0.2,0.2)^\top$, and covariance matrix $\text{cov}(\hm{y}_i\mid\hm{X}_i)=\phi\hm{\Sigma}_i=\phi \bm{A}_i^{1/2}\bm{R}(\alpha_y)\bm{A}_i^{1/2}$, $i=1,\dots,N_B$, where four covariates $\hm{x}_{ij[2:5]}\overset{iid}{\sim} \mathcal{N}_4(\hm{0},\hm{R}(\alpha_x))$ and intercept $\hm{x}_{ij[1]}=1$, $j=1,\dots,m$. Here both correlation matrices $\hm{R}(\alpha_x)$ and $\hm{R}(\alpha_y)$ are set as compound symmetry with $\alpha_x=0.5$ and $\alpha_y=0.7$, respectively. The dispersion parameter $\phi=1$ and the cluster size is $m=5$. We consider both marginal linear model for continuous ${y}_{ij}$ with $h(\mu_{ij})=\mu_{ij}$ and marginal logistic model for binary ${y}_{ij}$ with $h(\mu_{ij})=\log(\mu_{ij}/(1-\mu_{ij}))$. For all four methods in comparison, the working correlation matrix is specified to be compound symmetry which is also the true correlation structure.

\subsection{Evaluation of Parameter Estimation}
\noindent{\bf Scenario 1: fixed $N_B$ but varying batch size $n_b$}

\noindent We begin with the comparison of the four methods for the effect of data batch size $n_b$ on their point estimation and computational efficiency. A collection of $B$ data batches specified in Section~\ref{ssec:setup_QIF} are generated, each with data batch size $n_b$ and a total of $N_B=|D_B^\star|=10^5$ independent clusters, from an $m$-variate Gaussian linear model and an $m$-dimensional logistic model (using R package~\texttt{SimCorMultRes}). Tables~\ref{tab:linear1} reports the results of both linear and logistic MGLMs, over $500$ rounds of simulations. 

\noindent{\em Bias and coverage probability.} In linear and logistic MGLMs, Table~\ref{tab:linear1} shows that both RenewGEE and RenewQIF have similar bias and coverage probability in comparison to the two offline methods. This confirms the theoretical results given in Theorem~\ref{thm:normal_QIF23}; the RenewQIF as well as RenewGEE are stochastically equivalent to the offline QIF and offline GEE, respectively. It is easy to see that both bias and coverage probability in both the linear and logistic models are not affected by individual data batch size $n_b$. In other words, their performances seem to depend only on $N_B$.

\noindent{\em Computation time.} Two metrics are used to evaluate computation efficiency: ``C.Time" in Table~\ref{tab:linear1} refers to the total amount of time required by data loading and algorithm execution. With an increased $B$, both RenewGEE and RenewQIF show clearly advantageous for much lower computation time over the offline GEE or offline QIF, due to the fact that the two offline methods are much more time consuming to load in full datasets.\\

\noindent{\bf Scenario 2: fixed batch size $n_b$ but varying $B$}

\noindent Now we turn to a streaming setting where $B$ data batches arrive sequentially. For convenience, we fix single batch size $n_b=100$, but let $N_B$ increase from $10^3$ to $10^6$ (or $B$ from $10$ to $10^4$). Tables~\ref{tab:linear2} and~\ref{tab:logistic2} list the simulation results under the linear and logistic MGLMs.

\noindent{\em Bias and coverage probability.} As the number of batches $B$ rises from $10$ to $10^4$, both RenewGEE and RenewQIF confirm the asymptotic properties in Theorem~\ref{thm:normal_QIF1}: their average absolute bias decreases rapidly as the cumulative sample size accumulates, and the coverage probability stays robustly around the nominal level $95\%$. 

\noindent{\em Computation time.} Both online RenewGEE and RenewQIF methods show more and more advantageous as $N_B$ increases: the combined amount of time for data loading and algorithm execution only takes less than 5 seconds, whereas the offline GEE and offline QIF, when processing a dataset of $10^5$ clusters once, requires more than 20 seconds. This gain of 5-fold faster computation by the proposed RenewGEE and RenewQIF methods sacrifice little price of estimation precision or inferential power. One thing worth mentioning for Table~\ref{tab:logistic2} is that when $N_B=10^6$, to run the logistic MGLM, the offline GEE is computationally too intensive to produce convergent results within 12 hours using the existing R package~\texttt{gee}.

\begin{table}
	\caption{\label{tab:linear1}Simulation results under the linear and logistic MGLMs are summarized over $500$ replications, with fixed $N_B=10^5$ and $p=5$ with increasing number of data batches $B$. ``A.bias", ``ASE", ``ESE" and ``CP" stand for the mean absolute bias, the mean asymptotic standard error of the estimates, the empirical standard error, and the coverage probability, respectively. %``A.bias$\times10^{-3}$" indicates the scale of number in the cell, e.g. $1.10\times10^{-3}=0.0011$. 
		``C.Time" and ``R.Time" respectively denote computation time and running time, and the unit of both is second.}
	\begin{center}
		\small
		\begin{tabular}{l ccc ccc ccc ccc}
			\hline
			\hline
			&\multicolumn{12}{c}{\bf Linear MGLM} \\
			&\multicolumn{3}{c}{offline GEE} &\multicolumn{3}{c}{RenewGEE} &\multicolumn{3}{c}{offline QIF} &\multicolumn{3}{c}{RenewQIF}	   \\
			$B$ &100  &500  &2000  &100  &500  &2000  &100  &500  &2000  &100  &500  &2000\\
			\hline
			A.bias$\times 10^{-3}$ 
			&1.10  &1.10  &1.10
			&1.10  &1.10  &1.10
			&1.10  &1.10  &1.10
			&1.10  &1.10  &1.10\\
			ASE$\times 10^{-3}$  
			&1.42  &1.42  &1.42
			&1.42  &1.42  &1.42
			&1.42  &1.42  &1.42
			&1.42  &1.42  &1.42\\
			ESE$\times 10^{-3}$  
			&1.40  &1.40  &1.40
			&1.40  &1.40  &1.40
			&1.40  &1.40  &1.40
			&1.40  &1.40  &1.40\\ 
			CP  
			&0.95  &0.95  &0.95
			&0.95  &0.95  &0.95
			&0.95  &0.95  &0.95
			&0.95  &0.95  &0.95\\
			C.Time(s) 
			&9.27  &12.89  &20.56
			&2.70  &4.08   &8.45
			&2.60  &5.36   &13.91
			&1.39  &2.77   &6.63\\
			R.Time(s)         
			&8.53  &9.49  &8.53
			&2.31  &2.64  &3.62
			&1.86  &1.96  &1.88
			&1.06  &1.65  &2.95\\  
			\hline
			\hline
			&\multicolumn{12}{c}{\bf Logistic MGLM} \\
			&\multicolumn{3}{c}{offline GEE} &\multicolumn{3}{c}{RenewGEE} &\multicolumn{3}{c}{offline QIF} &\multicolumn{3}{c}{RenewQIF}	   \\
			$B$ &100  &500  &2000  &100  &500  &2000  &100  &500  &2000  &100  &500  &2000\\
			\hline
			A.bias$\times 10^{-3}$ 
			&2.70  &2.70  &2.70  
			&2.70  &2.70  &2.70
			&2.70  &2.70  &2.70
			&2.70  &2.70  &2.70 \\
			ASE$\times 10^{-3}$  
			&3.31  &3.31  &3.31
			&3.31  &3.31  &3.31
			&3.31  &3.31  &3.31
			&3.31  &3.31  &3.31  \\
			ESE$\times 10^{-3}$  
			&3.37  &3.37  &3.37
			&3.37  &3.37  &3.37
			&3.37  &3.37  &3.37
			&3.37  &3.37  &3.37  \\ 
			CP  
			&0.94  &0.94  &0.94
			&0.94  &0.94  &0.94
			&0.94  &0.94  &0.94
			&0.94  &0.94  &0.94 \\
			C.Time(s) 
			&10.73  &14.04  &27.77
			&2.05   &2.45  &3.32
			&3.22   &6.51  &20.25
			&1.14    &1.47  &2.37 \\
			R.Time(s)         
			&9.85   &9.86  &9.81
			&1.86    &2.07  &2.35
			&2.34   &2.33  &2.30 
			&0.99   &1.23  &1.86\\  
			\hline
		\end{tabular}
	\end{center}
\end{table}

\begin{table}
	\caption{\label{tab:linear2}Compare renewable estimators and offline ones in the linear MGLM model with fixed single batch size $n_b=100$ and $p=5$, $B$ increases from $10$ to $10^4$.}
	\begin{center}
		\small
		\begin{tabular}{l cccc| cccc}
			\hline
			\hline
			&\multicolumn{4}{c|}{$B=10$, $N_B=10^3$} 
			&\multicolumn{4}{c}{$B=100$, $N_B=10^4$}\\
		    &\multicolumn{2}{c}{GEE} &\multicolumn{2}{c|}{QIF}
			&\multicolumn{2}{c}{GEE} &\multicolumn{2}{c}{QIF} \\
			Criterion &offline &Renew &offline &Renew
			&offline &Renew &offline &Renew  \\
			\hline
			A.bias$\times 10^{-3}$ &11.06  &11.06  &11.08  &11.08
			&3.64  &3.64  &3.64  &3.64 \\
			ASE$\times 10^{-3}$  &14.19  &14.16  &14.15  &14.13
			&4.49  &4.49  &4.49  &4.49 \\
			ESE$\times 10^{-3}$  &13.82  &13.83  &13.85  &13.85
			&4.51  &4.51  &4.51  &4.51 \\
			CP  &0.956  &0.955  &0.952  &0.952
			&0.949  &0.947  &0.946  &0.946 \\	
			C.Time(s)  &0.033  &0.028  &0.019  &0.023
			&0.69  &0.25  &0.28  &0.18 \\
			R.Time(s)  &0.030  &0.024  &0.015  &0.019
			&0.58  &0.25  &0.16  &0.14 \\
			\hline
		   &\multicolumn{4}{c|}{$B=10^3$, $N_B=10^5$} 
		   &\multicolumn{4}{c}{$B=10^4$, $N_B=10^6$}\\
		   &\multicolumn{2}{c}{GEE} &\multicolumn{2}{c|}{QIF}
		   &\multicolumn{2}{c}{GEE} &\multicolumn{2}{c}{QIF} \\
		   Criterion &offline &Renew &offline &Renew
		   &offline &Renew &offline &Renew  \\
		   \hline
		   A.bias$\times 10^{-3}$ &1.11  &1.11  &1.11  &1.11
		   &0.35  &0.35  &0.35  &0.35  \\
		  ASE$\times 10^{-3}$  &1.42  &1.42  &1.42  &1.42
		  &0.45  &0.45  &0.45  &0.45   \\
		  ESE$\times 10^{-3}$  &1.40  &1.40  &1.40  &1.40
		  &0.44  &0.44  &0.44  &0.44 \\
		  CP  &0.952  &0.954  &0.952  &0.952
		  &0.955  &0.955  &0.955  &0.955 \\
		  C.Time(s)  &15.38  &5.70  &8.57  &4.26
		  &781.46  &62.30  &704.11  &51.38 \\
		  R.Time(s)  &8.72  &2.96  &1.90  &2.18
		  &99.21  &32.12  &21.85  &25.21 \\
			\hline
		\end{tabular}
	\end{center}
\end{table}

\begin{table}
	\caption{Compare renewable estimators and offline ones in the logistic MGLM model with fixed single batch size $n_b=100$ and $p=5$, $B$ increases from $10$ to $10^4$. The dashed line in the column for ``offline GEE" when $N_B=10^6$ indicates the standard~\texttt{gee} package in R does not produce output due to the excessive computational burden.}
	\begin{center}
		\small
		\begin{tabular}{l cccc| cccc}
				\hline
				\hline
			&\multicolumn{4}{c|}{$B=10$, $N_B=10^3$} 
			&\multicolumn{4}{c}{$B=100$, $N_B=10^4$}\\
			&\multicolumn{2}{c}{GEE} &\multicolumn{2}{c|}{QIF}
			&\multicolumn{2}{c}{GEE} &\multicolumn{2}{c}{QIF} \\
			Criterion &offline &Renew &offline &Renew
			&offline &Renew &offline &Renew  \\
			\hline
			A.bias$\times 10^{-3}$ &25.92  &25.82  &26.07  &26.01
			&8.17  &8.16  &8.17  &8.16\\
			ASE$\times 10^{-3}$  &33.08  &33.06  &33.03  &33.07
			&10.45  &10.45  &10.45  &10.45\\
			ESE$\times 10^{-3}$  &32.48  &32.36  &32.67  &32.60
			&10.31  &10.30  &10.32  &10.31\\
			CP  &0.953  &0.952  &0.950  &0.952
			&0.950  &0.952  &0.951  &0.952  \\	
			C.Time(s)  &0.048  &0.029  &0.024  &0.023
			&1.09  &0.23  &0.32  &0.17 \\
			R.Time(s)  &0.045  &0.026  &0.021  &0.020
			&0.99  &0.20  &0.22  &0.14\\
			\hline			
			&\multicolumn{4}{c|}{$B=10^3$, $N_B=10^5$} 
			&\multicolumn{4}{c}{$B=10^4$, $N_B=10^6$}\\
			&\multicolumn{2}{c}{GEE} &\multicolumn{2}{c|}{QIF}
			&\multicolumn{2}{c}{GEE} &\multicolumn{2}{c}{QIF} \\
			Criterion &offline &Renew &offline &Renew
			&offline &Renew &offline &Renew  \\
		 \hline
		A.bias$\times 10^{-3}$ &2.71  &2.71  &2.71  &2.71
		&-  &0.82  &0.82  &0.82\\
		ASE$\times 10^{-3}$  &3.31  &3.31  &3.31  &3.31 
		&-  &1.05  &1.05  &1.05\\
		ESE$\times 10^{-3}$  &3.39  &3.39  &3.39  &3.39
		&-  &1.04  &1.04  &1.04\\
		CP  &0.948  &0.948  &0.948  &0.948 
		&-  &0.946  &0.946  &0.948\\
		C.Time(s)  &22.41  &5.84  &9.99  &4.49 
		&-  &57.47  &856.70  &47.44\\
		R.Time(s)  &15.59  &3.02  &3.18  &2.35 
		&-  &31.08  &45.83  &21.31\\
		\hline
		\end{tabular}
		\label{tab:logistic2}
	\end{center}
\end{table}

\subsection{Evaluation of Monitoring Procedure}
We also evaluate the performance of the proposed diagnostic procedure using the goodness-of-fit test $\Lambda_b$ in equation~\eqref{eq:test} to detect abnormal data batches. First, we check the properties of this test statistic with respect to both type I error and power of detection abnormal data batches. Then, we compare the estimation and inference performance of the RenewQIF methods with and without the use of monitoring procedure in terms of the following four criterion: (a) A.bias, (b) ASE, (c) ESE, and (d) CP, as define above. The abnormal data batches are created by altering the true parameters via a local departure on $\beta_{02}$, that is $\hm{\beta}_\tau=(0.2, -(0.2+d), 0.2, -0.2, 0.2)^\top$, $\tau\in\Gamma_q$. We set $\Gamma_2$, containing two positions ($q=2$) at which two abnormal data batches occurs, respectively, at $\tau_1=0.25B$ and $\tau_2=0.75B$. Simulation studies have showed that type I errors are very close to the nominal level $\alpha$, and that the power of detecting abnormal data batches drops as $\alpha$ becomes smaller, but increases as $n_b$ increases. See Table 1 in the Supplementary Material.
%Table~\ref{tab:empirical} shows the empirical type I error rates for different batch sizes $n_b$ under various significance level $\alpha\in\{0.1, 0.05, 0.01, 0.001, 5\times10^{-6} \}$. They are all very close to the nominal level $\alpha$. These findings confirm the theoretical insights for a fixed local alternative with departure size $d$. Table~\ref{tab:empirical} also shows that the power of detection abnormal data batches drops as $\alpha$ becomes smaller, while the power increases with increasing $n_b$.

\noindent{\bf Without monitoring procedure.} 
With fixed $N_B=10^4$ and $\Gamma_2=\{0.25B, 0.75B \}$, the upper panel in Table~\ref{tab:fixN_qc} shows that larger data batch size $n_b$ leads to a larger bias due to the increased number of contaminated clusters generated from the incompatible data model. A.bias increases almost linearly with $n_b$. At similar levels of ASE and ESE, the coverage probability deviates more from the nominal level $95\%$ as $n_b$ increases; it drops from $93.2\%$ to $3.0\%$ as $n_b$ rises from $50$ to $200$ due to more severe data contamination.

\noindent{\bf With monitoring procedure.} For the purpose of quality control, larger $\alpha$ increases the sensitivity of rejection, so many small departures may be detected, which would be consequently ignored in the online updating. These are clearly shown in the lower panel of Table~\ref{tab:fixN_qc} due to the reduced proportion of used samples, defined by $N_0/N_B$. See also the last subplot in Figure 1 in the Supplementary Material where $N_0$ is the number of clusters that passed the diagnostic test. The price to pay in this case is that the resulting bias and standard error would be larger than they would be if the false positives may be avoided. In contrast, choosing a small $\alpha$ may elevate type II error (false negative) and thus can lose power in detecting abnormal data batches. In this case, the price to pay is not only increased bias but also decreased coverage probability. The latter is indeed a more serious problem as far as inference concerns. This phenomenon is evident when $n_b$ is small as shown in Table~\ref{tab:fixN_qc}. As an extreme case of $\alpha=5\times10^{-6}$, the detection power is greatly lost with $n_b=200$, $100$ or $50$, and the coverage probability reduces to lower than $90\%$. In practice, with high throughput data streams, where cumulative sample sizes increase rapidly, using a larger $\alpha$, say $\alpha=0.05$, is much safer and recommended in practice for the purpose of monitoring, resulting in a more protective scenario by effectively and cautiously avoiding abnormal data batches.

\begin{table}[h]
	\caption{\label{tab:fixN_qc}Performances with and without monitoring procedure. Fixed total number of samples $N_B=10^4$ with varying data batch size $n_b$. $\tau_1=0.25B$ and $\tau_2=0.75B$. In the table ``With monitoring procedure", $N_0/N_B$ denotes the proportion of used samples in the renewable estimation and inference.}
	\begin{center}
		\small
		\begin{tabular}{l l l  l l}
			\hline
			\hline
			&\multicolumn{4}{c}{\bf Without monitoring procedure}  \\
			\hline
			$n_b$ &50 &100 &200 &400 \\
			\hline
			A.bias$\times 10^{-3}$    
			&12.36  &16.74  &28.68  &56.58 \\
			ASE$\times 10^{-3}$     
			&14.61  &14.62  &14.69  &14.79 \\
			ESE$\times 10^{-3}$    
			&14.25  &14.02  &15.26  &15.03 \\
			CP  &0.932   &0.838   &0.534   &0.030   \\
			\hline
		\end{tabular}
	\end{center}
\vspace{-0.8cm}
	\begin{center}
		\small
		\begin{tabular}{l   l l l l l | l l l l l}
			\hline
			\hline
			&\multicolumn{10}{c}{\bf With monitoring procedure}  \\
			&\multicolumn{5}{c|}{$n_b=50$} &\multicolumn{5}{c}{$n_b=100$} \\
			\hline
			$\alpha\times 10^{-3}$  &100  &50 &10 &1  &0.005
			&100  &50 &10 &1  &0.005 \\
			\hline
			A.bias$\times 10^{-3}$  
			&\bf11.29 &8.923 &7.970 &8.312  &8.802
			&\bf9.773 &8.433 &7.962 &8.147  &\bf9.564 \\
			ASE$\times 10^{-3}$    
			&10.18 &9.558 &9.259  &9.220  &9.219
			&10.21  &9.604 &9.311 &9.263  &9.256 \\
			ESE$\times 10^{-3}$     
			&20.90 &12.08 &9.860 &9.782  &9.568
			&16.67 &11.37 &9.834 &9.340  &9.736 \\
			CP                                     
			&\bf0.852 &\bf0.896 &0.920 &0.906  &\bf0.890
			&\bf0.926 &0.942 &0.942 &0.940  &\bf0.882  \\
			$N_0/N_B$  
			&0.876  &0.939  &0.988  &0.997  &0.999 
			&0.883 &0.933  &0.976  &0.987  &0.992 \\
			\hline
			
			&\multicolumn{5}{c|}{$n_b=200$ } &\multicolumn{5}{c}{$n_b=400$} \\
			\hline
			$\alpha\times 10^{-3}$   &100  &50 &10 &1  &0.005
			&100  &50 &10 &1  &0.005 \\
			\hline
			A.bias$\times 10^{-3}$   
			&8.836   &8.122  &7.485  &7.757   &\bf9.641
			&9.317  &8.371  &7.723  &7.656  &7.771  \\
			ASE$\times 10^{-3}$     
			&10.16  &9.709  &9.426  &9.369   &9.347	
			&10.39  &9.958  &9.638  &9.584  &9.573 \\
			ESE$\times 10^{-3}$     
			&11.73   &10.41 &9.305  &9.551   &10.40
			&12.41   &10.79  &9.690  &9.612   &9.795 \\
			CP           
			&0.938  &0.948  &0.956  &0.946   &\bf0.886
			&0.934   &0.948  &0.958  &0.956  &0.952 \\
			$N_0/N_B$ 
			&0.863  &0.913  &0.952  &0.962  &0.972
			&0.823  &0.872  &0.911  &0.918  &0.920 \\
			\hline
		\end{tabular}
	\end{center}
\end{table}

\clearpage
\section{Analysis of NASS CDS Data}\label{sec:realdata}
In regard to injuries involved in car accidents, we are interested in not only the extent of injuries in drivers but also in passengers. Apparently, injury levels of driver and passengers within the same vehicle are correlated, and such within-cluster correlation needs to be taken into account in the analysis. In this real data application, we focus on the analysis of a series of car crash datasets from the National Automotive Sampling System-Crashworthiness Data System (NASS CDS) from January, 2009 to December, 2015. Our primary interest was to evaluate the effectiveness of graduated driver licensing (GDL) on overall driving safety with respect to injury levels in both driver and passengers. GDL is a nationwide legislature on novice drivers of age 21 or younger with various conditions of vehicle operation. In contrast, under the current law, there are no restrictions on vehicle operation for older drivers with age, {\em say}, older than 65. Thus, we want to compare drivers' age groups with respect to the extent of injury when a car accident happens. We first categorized the ``Age" variable into three age groups: ``Age$<$21" represents the young group under a restricted GDL, and ``Age$\geq$65" indicates the old group with a regular full driver's license, while those of age in between is treated as the reference group. 
%Three age groups (Age$<$21, 21$\leq$Age$<$65, Age$\geq$65) were coded as dummy variables in our analysis. Since the number of drivers and passengers involved in accidents in either young or old age group was much smaller than those in the reference group, it was of great interest to update analysis results with more data being collected sequentially over time. 
Extent of injury in a crash is a binary variable with 1 for a moderate or severe injury, and 0 for minor or no injury. This outcome variable was created from the variable of Maximum Known Occupant Ais (MAIS), which indicates the single most severe injury level reported for each occupant. Other potential risk factors are also considered in the model, including seat belt use (Seat Belt, 1 for used and 0 for no), drinking (Drinking, 1 for yes and 0 for no), speed limit (Speed Limit), vehicle weight (Vehicle Weight, 0 for $\leq3000$, 1 for 3000$\sim$4000, 2 for $\geq$4000 ), air bag system deployed (Air Bag, 1 for yes and 0 for no), number of lanes (Number of Lanes, 0 for $\leq2$ and 1 for else), drug involvement in this accident (Drug Use, 1 for yes and 0 for no), driver's distraction/inattention to driving (Distraction, 1 for attentive and 0 for else), roadway surface condition (Surface Condition, 1 for dry and 0 for else), and has vehicle been in previous accidents (Previous Accidents, 0 for no and 1 for else). 

Streaming data are formed by quarterly accident data from the period of $7$ years from January, 2009 to December 2015, with $B=28$ data batches and a total of $N_B=18,832$ crashed vehicles that contain $26,330$ occupants with complete records. Each vehicle is treated as a cluster, and the cluster size varies from 1 to 10 with an average of 2 occupants. We invoke RenewQIF to fit a marginal logistic regression model with the compound symmetry correlation to account for the within vehicle correlation. In the analysis, we are interested in a 7-year average risk assessment and thus assuming constant associations between extent of injuries and risk factors over time. Since this sequence of data streams arrive at low speed and large data batch size, we would expect to have high power to detect the abnormal data batch even if we choose a small $\alpha$, say $\alpha=0.01$. Additionally, since samples from NASS CDS have gone through extensive data cleaning and pre-processing steps, a stringent $\alpha$ is a reasonable choice to make a full use of samples. At $\alpha=0.01$, our proposed monitoring procedure identifies data batch $\mathcal{D}_8$ to be incompatible, corresponding to the data collected during the 4th quarter in year 2010. Estimated coefficients, standard errors and $p$-values are reported in Table 2 in the Supplementary Material.

Figure~\ref{fig:CDS_p_QIF} shows the trajectories of $-\log_{10}(p)$ values of the online Wald test developed for RenewQIF, each for one regression coefficient over $28$ quarters. Even though the total sample size $N_B$ increases over time, not all of them show steep monotonic increasing trends in evidence against the null $H_0: \beta_j=0$. ``Seat Belt" turns out to have the strongest association to the odds of injury in a crash among all covariates included in the model. This is an overwhelming confirmation to the enforcement of policy ``buckle up" when sitting in a moving vehicle. For the convenience of comparison, we report a summary statistic as of the area under the $p$-value curve for each covariate. ``Seat Belt" (2297.45), ``Drug Use" (1779.85), ``Air Bag" (1249.49) and ``Previous Accidents" (1342.46) appear well separated from the other risk factors. Their ranking is well aligned with the ranking of $p$-values obtained at the end time of streaming data availability, namely December, 2015. 

The trajectories of both young and old age groups are evaluated in Figure~\ref{fig:compare_beta_QIF} with or without data batch $\mathcal{D}_8$. The trace of the estimators for the young age group (Age$<$21) stays below 0 over the 28-quarter period, indicating that it has lower adjusted odds of moderate/severe injury than the reference age group. This finding confirms the effectiveness of GDL in protecting young novice drivers. Unfortunately, in contrast, the old age group (Age$\geq$65) turns out to suffer from significantly higher adjusted odds of moderate/severe injury outcomes comparing to the middle age group. This suggests a need of policy-making to protect older drivers from injuries when an accident happens. Furthermore, the abnormal data batch seems to affect marginally the estimates for age groups if we compare the plots with (right) and without (left) monitoring procedure (see the red vertical dashed line).

Applying the proposed RenewQIF to the above CDS data analysis enables us to visualize time-course patterns of data evidence accrual as well as stability and reproducibility of inference. As shown in Figure~\ref{fig:compare_beta_QIF}, at the early stage of data streams, due to limited sample sizes and possibly sampling bias, both parameter estimates and test power may be unstable and even possibly misleading. These potential shortcomings can be overcome when estimates and inferential quantities are continuously updated along with data streams, which eventually reach stability and reliable conclusions.

\begin{figure}[h]
	\begin{center}
		\includegraphics[width=0.8\linewidth]{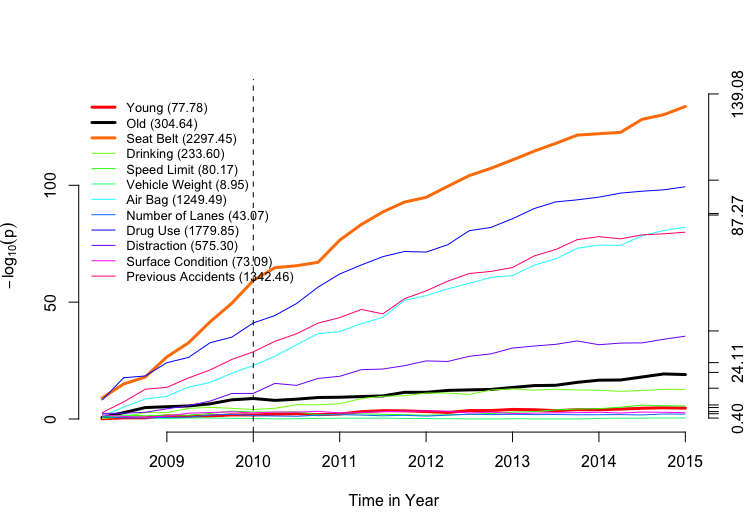}
		\vspace{-0.8cm}
		\caption{\small Trace plots of $-\log_{10}(p)$ over quarterly data batches from January, 2009 to December, 2015, each for one regression coefficient. Dashed vertical line indicates the location of detected abnormal data batch.}
		\label{fig:CDS_p_QIF}
	\end{center}
\end{figure}

\begin{figure}[h]
	\begin{center}
		\includegraphics[width=0.7\linewidth]{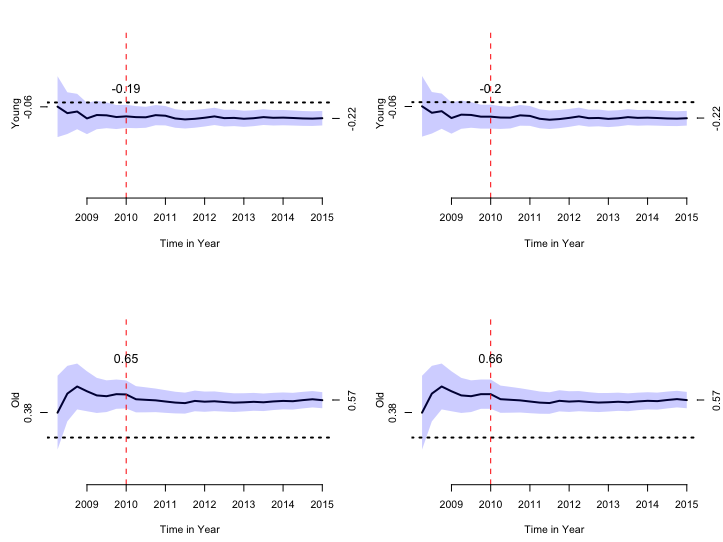}
		\caption{\small Trace plots for the coefficients estimates and $95\%$ pointwise confidence bands of ``Young" and ``Old". Numerical numbers on two sides denote the estimated regression coefficients after the arrival of first and last batches, while the ones above the traces denote the estimates at the $8$-th data batch.}
		\label{fig:compare_beta_QIF}
	\end{center}
\end{figure}

\section{Concluding Remarks}
Due to the advantage of technologies in data storage and data collection, streaming data arise from many practical areas such as healthcare~\citep{Peek2014}. Healthcare data are typically measurements in forms of clusters or time series, such as patients from the same clinic, or data from personal wearable devices~\citep{Wu2016}. The traditional methods for clustered/longitudinal data analysis such as generalized estimating equations (GEE) and quadratic inference functions (QIF) that process the entire dataset once may be greatly challenged due to the following reasons: (i) they become computationally prohibitive as the total sample size accumulates too fast and too large, so to exceed the available computational power; see Table~\ref{tab:logistic2} where GEE fails to produce numerical outputs; and (ii) historical subject-level data may no longer be accessible due to storage, time, or privacy issues. This type of problem has been extensively tackled in the framework of online updating where stochastic gradient descent algorithms are the primary methods of choice to provide fast updating with no use of historical data. However, most online learning algorithms have not considered statistical inference or the diagnosis of contaminated data batches. 

Such gaps have been filled in this paper by the RenewQIF methodology. The proposed RenewQIF method provides a new paradigm of renewable estimation and incremental inference, in which parameter estimates are recursively updated with current data and inferential statistics of historical data, but does not require the accessibility to any historical subject-level data. To achieve efficient communications between current data and historical summary statistics, we design an extended Spark's Lambda architecture to execute both data storage and analysis updates. Both proposed statistical methodology and computational algorithms have been investigated for their theoretical guarantees and examined numerically via extensive simulation studies. The proposed RenewQIF has been shown to be much more computationally efficient with no loss of inferential power in comparison to the offline GEE or offline QIF. Additionally, we propose a diagnostic procedure to detect abnormal data batches in data streams for the proposed RenewQIF. The utilization of a goodness-of-fit test in the QIF framework enabled us to check the compatibility of current data batch with a normal reference effectively and efficiently. The proposed monitoring procedure has been integrated into an extended Lambda architecture with an additional monitoring layer.

The formulation of RenewQIF is under the assumption that clusters arrive independently over data streams, and when cluster size $m=1$, it reduces to generalized linear model as a special case. A direction of interest is to consider the case of inter-correlated batches, such as serially dependent data streams generated by individual wearable devices. Such types of data streams are pervasive in healthcare where thousands of physiological measurements are recorded per second, such as body temperature, heart rate, respiratory rate and blood pressure~\citep{Healthcare2014}. Therefore, developing the analytic tools for the analysis of serially dependent data streams is an important future research as part of new analytics for handling massive data volumes and making behavioral interventions.

%\bigskip
%\begin{center}
%	{\large\bf SUPPLEMENTARY MATERIAL}
%\end{center}
\appendix

\setcounter{equation}{0}
\makeatletter
\renewcommand{\theequation}{a.\arabic{equation}}
\renewcommand{\thesubsection}{A.\arabic{subsection}}
\section*{Appendix}

\begin{lemma}\label{pf:lemma}
	Recall that $n_j$ and $N_j$ are the batch size of $\mD_j$ and the cumulative sample size up to $j$, respectively, $j=1,\dots,b$. Then,
	\begin{equation}\label{eq:lemma11}
	\sum_{j=1}^{b}\frac{n_j}{N_j} \leq 1 + \log\frac{N_b}{n_1}, 
	\end{equation}
	\begin{equation}\label{eq:lemma12}
	\sum_{j=1}^{b} \frac{n_j}{\sqrt{N_j}} \leq 2\sqrt{N_b}.
	\end{equation}
\end{lemma}
Proof: We first prove~\eqref{eq:lemma11}. Let $f(x)=\log(1+x)-\frac{x}{1+x}$, $x>0$. Since $f(0)=0$ and $f'(x)>0$ for $x>0$, we have $f(x)\geq 0$. Let $x=\frac{n_b}{N_{b-1}}$, and it follows that $\log\left(N_b/N_{b-1}\right) > n_b/N_b$, namely, $\log(N_b)\geq \log(N_{b-1})+n_b/N_b$. Repeat the above procedure and let $x=n_j/N_{j-1}$ for $j=b-1,\dots,2$:
\[
\begin{split}
\log (N_b) &\geq \log(N_{b-1}) + \frac{n_b}{N_b} \\
&\geq \log(N_{b-2}) + \frac{n_{b-1}}{N_{b-1}} + \frac{n_b}{N_b} \\
\dots & \geq \log(N_1) + \sum_{j=2}^{b}\frac{n_j}{N_j}.
\end{split}
\]

Now we prove~\eqref{eq:lemma12}. First, note that fact that 
\[
2(\sqrt{m+q} - \sqrt{m}) \geq \frac{q}{\sqrt{m+q}}, \ \text{for} \ m,q > 0.
\]
Let $m = N_{b-1}$ and $q=n_b$, then we have $2\sqrt{N_b}\geq 2\sqrt{N_{b-1}} + n_b / \sqrt{N_b}$. Now we repeat this procedure by setting $m = N_{j-1}$, $q = n_j$ for $j=b-1,\dots, 2$. It follows that
\[
\begin{split}
2 \sqrt{N_b} & \geq 2\sqrt{N_{b-1}} + \frac{n_b}{\sqrt{N_b}} \\
& \geq 2\sqrt{N_{b-2}} + \frac{n_{b-1}}{\sqrt{N_{b-1}}} + \frac{n_b}{\sqrt{N_b}} \\
\dots & \geq 2\sqrt{N_1} + \sum_{j=2}^{b} \frac{n_j}{\sqrt{N_j}}.
\end{split}
\]
Since $n_1 = N_1$, we finish the proof of this lemma.

\subsection{Large Sample Properties in Scenarios 2 and 3}\label{app:pf_QIF2}
The quadratic inference function estimator of the cumulative dataset to time point $b$ is $\hm{\hat{\beta}}_b^\star =\underset{\hm{\beta}\in\mathbb{R}^p}{\arg \text{min}}\ Q_b^\star(\hm{\beta})$. Let $\hm{\beta}_0$ be the true parameter and $\hm{\tilde{\beta}}_b$ be the renewable estimator. Note that for the first data batch $\mathcal{D}_1$, we have $\hm{\tilde{\beta}}_1=\hm{\hat{\beta}}_1^\star=\hm{\hat{\beta}}_1$, which is consistent by the existing theory of the generalized method of moments (GMM) estimators~\citep{Hansen1982}. Now we prove the consistency of $\bm{\tilde{\beta}}_b$ for an arbitrary $b\geq2$ by the method of induction.

We have defined an estimating function
\begin{equation*}
\begin{split}
\bm{f}_b(\hm{\beta})&=\frac{1}{N_b}\left\{\sum_{j=1}^{b-1}\hm{G}_j(\hm{\tilde{\beta}}_j)+\hm{G}_b(\hm{\beta}) \right\}^\top \left\{\sum_{j=1}^{b-1}\hm{C}_j(\hm{\tilde{\beta}}_j)+\hm{C}_b(\hm{\beta}) \right\}^{-1}\\ &\times\left\{\sum_{j=1}^{b-1}\hm{g}_j(\hm{\tilde{\beta}}_j)+\sum_{j=1}^{b-1}\hm{G}_j(\hm{\tilde{\beta}}_j)(\hm{\tilde{\beta}}_{b-1}-\hm{\beta})+\hm{g}_b(\hm{\beta}) \right\},
\end{split}
\end{equation*}
and the renewable estimator $\hm{\tilde{\beta}}_b$ satisfies:
$\bm{f}_b(\hm{\tilde{\beta}}_b)=\hm{0}$.

%Since $\hm{\tilde{\beta}}_{j}$'s satisfy $P(\sqrt{N_j}\|\bm{\tilde{\beta}}_j-\bm{\beta}_0\|>\eta)\leq C_\eta$ for $j=1,\dots,b-1$, we have
\begin{equation}\label{pfeq:f_beta0}
\begin{split}
\bm{f}_b(\hm{\beta}_0)&=\frac{1}{N_b}\left\{\sum_{j=1}^{b-1}\hm{G}_j(\hm{\tilde{\beta}}_j)+\hm{G}_b(\hm{\beta}_0) \right\}^\top \left\{\sum_{j=1}^{b-1}\hm{C}_j(\hm{\tilde{\beta}}_j)+\hm{C}_b(\hm{\beta}_0) \right\}^{-1} \\
& \times \left\{\sum_{j=1}^{b-1}\hm{g}_j(\hm{\tilde{\beta}}_j)+\sum_{j=1}^{b-1}\hm{G}_j(\hm{\tilde{\beta}}_j)(\hm{\tilde{\beta}}_{b-1}-\hm{\beta}_0) + \hm{g}_b(\hm{\beta}_0) \right\} \\
&= \left\{\frac{1}{N_b}\sum_{j=1}^{b}\hm{G}_j(\hm{\beta}_0) +\Delta_1 \right\}^\top\left\{\frac{1}{N_b}\sum_{j=1}^{b}\hm{C}_j(\hm{\beta}_0)+\Delta_2 \right\}^{-1} \\ &\times\left\{\frac{1}{N_b}\sum_{j=1}^{b}\hm{g}_j(\hm{\beta}_0)+\Delta_3\right\}, \\
%& + \frac{1}{N_b}\bm{\tilde{G}}_b(\bm{\beta}_0)^\top\left\{ \bm{\tilde{C}}_b(\bm{\beta}_0)+\Delta_2\right\}^{-1} \Delta_3+ \frac{1}{N_b} \Delta_1^\top  \left\{ \bm{\tilde{C}}_b(\bm{\beta}_0)+\Delta_2\right\}^{-1} \bm{\tilde{g}}_b(\bm{\beta}_0) \\
%& + \frac{1}{N_b} \Delta_1^\top \left\{ \bm{\tilde{C}}_b(\bm{\beta}_0)+\Delta_2\right\}^{-1} \Delta_3,
\end{split}
\end{equation}
where $\Delta_1=\frac{1}{N_b}\sum_{j=1}^{b-1}\left\{\bm{G}_j(\bm{\tilde{\beta}}_j) - \bm{G}_j(\bm{\beta}_0) \right\}$, $\Delta_2=\frac{1}{N_b}\sum_{j=1}^{b-1}\left\{\bm{C}_j(\bm{\tilde{\beta}}_j) - \bm{C}_j(\bm{\beta}_0) \right\}$, and \\ $\Delta_3=\frac{1}{N_b}\sum_{j=1}^{b-1}\hm{G}_j(\hm{\tilde{\beta}}_j)(\hm{\tilde{\beta}}_{b-1}-\hm{\beta}_0) +\frac{1}{N_b}\sum_{j=1}^{b-1}\left\{\bm{g}_j(\bm{\tilde{\beta}}_j) - \bm{g}_j(\bm{\beta}_0) \right\}$. Since $\bm{G}_j$ is Lipschitz continuous in $\Theta$, there exists $M_1(\mathcal{D}_j)>0$ such that $\|\bm{G}_j(\bm{\tilde{\beta}}_j) - \bm{G}_j(\bm{\beta}_0)\|\leq M_1(\mathcal{D}_j)\|\bm{\tilde{\beta}}_j-\bm{\beta}_0\|$. Additionally, by condition~\ref{C3:bound_G} that $\bm{g}_j$ is continuously differentiable, $\bm{C}_j$ is also continuous; and by condition~\ref{C4:bound_C}, $\bm{C}_j$ is bounded in $\Theta$, there also exists $M_2(\mathcal{D}_j)>0$ such that $\|\bm{C}_j(\bm{\tilde{\beta}}_j) - \bm{C}_j(\bm{\beta}_0)\|\leq M_2(\mathcal{D}_j)\|\bm{\tilde{\beta}}_j-\bm{\beta}_0\|$. Furthermore, according to mean value inequality~\citep{Rudin1976}, there exists $\bm{\xi}_j$ between $\bm{\tilde{\beta}}_j$ and $\bm{\beta}_0$ such that $\|\bm{g}_j(\bm{\tilde{\beta}}_j) - \bm{g}_j(\bm{\beta}_0)\|\leq\|\bm{G}_j(\bm{\xi}_j)(\bm{\tilde{\beta}}_j - \bm{\beta}_0) \|$ with $\bm{G}_j(\bm{\xi}_j)=O(n_j)$ according to condition~\ref{C3:bound_G}. Therefore, we can prove
\[
\Delta_1 = \frac{1}{N_b}\sum_{j=1}^{b-1}\left\{ \bm{G}_j(\bm{\tilde{\beta}}_j)-\bm{G}_j(\bm{\beta}_0)\right\} \leq \frac{1}{N_b}\sum_{j=1}^{b-1} M_1(\mathcal{D}_j)\|\bm{\tilde{\beta}}_j-\bm{\beta}_0 \|  =\frac{1}{N_b}O_p\left(\sum_{j=1}^{b-1}\frac{n_j}{\sqrt{N_j}} \right),
\]
\[\Delta_2 = \frac{1}{N_b}\sum_{j=1}^{b-1} \left\{\bm{C}_j(\bm{\tilde{\beta}}_j) - \bm{C}_j(\bm{\beta}_0)\right\} \leq \frac{1}{N_b} \sum_{j=1}^{b-1} M_2(\mathcal{D}_j)\|\bm{\tilde{\beta}}_j-\bm{\beta}_0\| = \frac{1}{N_b}O_p\left(\sum_{j=1}^{b-1}\frac{n_j}{\sqrt{N_j}} \right),\]
\[\Delta_3 \leq \frac{1}{N_b} O_p\left(\sum_{j=1}^{b-1}\frac{n_j}{\sqrt{N_{b-1}}} + \frac{1}{N_b} \sum_{j=1}^{b-1} \frac{n_j}{\sqrt{N_j}} \right).
\]
Note that according to Lemma 1, the summation $\sum_{j=1}^{b-1}\frac{n_j}{\sqrt{N_j}}\leq 2\sqrt{N_{b-1}}$, and therefore the term $\frac{1}{N_b}O_p\left(\sum_{j=1}^{b-1}\frac{n_j}{\sqrt{N_j}} \right)=o_p(1)$ as long as $N_b\to\infty$, i.e. $b$ can be either finite or $b\to\infty$.
In addition, by the property of QIF estimating function $\bm{\tilde{G}}_b(\bm{\beta}_0)^\top\bm{\tilde{C}}_b(\bm{\beta}_0)^{-1}\bm{\tilde{g}}_b(\bm{\beta}_0)=O_p(\sqrt{N_b})$, it follows that $\bm{f}_b(\bm{\beta}_0)=O_p(N_b^{-1/2})$. Taking a difference between $\bm{f}_b(\bm{\beta}_0)$ and $\bm{f}_b(\bm{\tilde{\beta}}_b)$, we get
\begin{equation}\label{pfeq:beta0-beta}
\begin{split}
\bm{f}_b(\hm{\beta}_0)-\bm{f}_b(\hm{\tilde{\beta}}_b)
&=\frac{1}{N_b}
\left\{\sum_{j=1}^{b-1}\hm{G}_j(\hm{\tilde{\beta}}_j)+\hm{G}_b(\hm{\beta}_0) \right\}^\top \left\{\sum_{j=1}^{b-1}\hm{C}_j(\hm{\tilde{\beta}}_j)+\hm{C}_b(\hm{\beta}_0) \right\}^{-1}\\
&\times\left\{\sum_{j=1}^{b-1}\hm{G}_j(\hm{\tilde{\beta}}_j)(\hm{\tilde{\beta}}_b-\hm{\beta}_0)+\hm{g}_b(\hm{\beta}_0)-\hm{g}_b(\hm{\tilde{\beta}}_b) \right\}\\
&-\frac{1}{N_b}
\left[ 
\left\{\sum_{j=1}^{b} \hm{G}_j(\hm{\tilde{\beta}}_j) \right\}^\top
\left\{\sum_{j=1}^{b} \hm{C}_j(\hm{\tilde{\beta}}_j) \right\}^{-1} 
\right.\\
&\left. 
-\left\{\sum_{j=1}^{b-1}\hm{G}_j(\hm{\tilde{\beta}}_j)+\hm{G}_b(\hm{\beta}_0) \right\}^\top\left\{\sum_{j=1}^{b-1}\hm{C}_j(\hm{\tilde{\beta}}_j)+\hm{C}_b(\hm{\beta}_0) \right\}^{-1} \right] \\
&\times
\left\{\sum_{j=1}^{b}\hm{g}_j(\hm{\tilde{\beta}}_j)
+\sum_{j=1}^{b-1}\hm{G}_j(\hm{\tilde{\beta}}_j)(\bm{\tilde{\beta}}_{b-1}-\bm{\tilde{\beta}}_b) \right\}.
\end{split}
\end{equation}

According to mean value inequality, since $\bm{g}(\cdot)$ is continuous in $\bm{\beta}\in\Theta$, there exists $\bm{\xi}_b$ between $\bm{\tilde{\beta}}_b$ and $\bm{\beta}_0$ such that $\|\bm{g}_b(\bm{\tilde{\beta}}_b) - \bm{g}_b(\bm{\beta}_0)\|\leq\|\bm{G}_b(\bm{\xi}_b)(\bm{\tilde{\beta}}_b-\bm{\beta}_0)\|$. Thus, we have
\begin{equation*}
	\begin{split}
	\bm{G}_b(\bm{\xi}_b)(\bm{\tilde{\beta}}_b-\bm{\beta}_0) &= 
	\left\{\hm{G}_b(\hm{\xi}_b)+ \hm{G}_b(\hm{{\beta}}_0)-\hm{G}_b(\hm{\beta}_0)\right\}(\hm{\tilde{\beta}}_b-\hm{\beta}_0) \\
	&=\hm{G}_b(\hm{{\beta}}_0)(\hm{\tilde{\beta}}_b-\hm{\beta}_0)+\left\{\hm{G}_b(\hm{\xi}_b) -\hm{G}_b(\hm{{\beta}}_0) \right\}(\hm{\tilde{\beta}}_b-\hm{\beta}_0).
	\end{split}
\end{equation*}
Since $\hm{G}_b=\sum_{i\in\mathcal{D}_b}\hm{G}(\hm{y}_i;\hm{X}_i,\hm{\beta})$ is Lipschitz continuous in $\Theta$, there exists $M_1(\mathcal{D}_b)>0$ such that $\|\hm{G}_b(\hm{{\beta}}_0)-\hm{G}_b(\hm{\xi}_b) \| \leq M(\mathcal{D}_b)\|\hm{\xi}_b-\hm{\beta}_0 \| \leq M_1(\mathcal{D}_b)\| \hm{\tilde{\beta}}_b-\hm{\beta}_0\|$, and it follows that 
\begin{equation}\label{pfeq:mean}
	\|\hm{g}_b(\hm{\tilde{\beta}}_b)-\hm{g}_b(\hm{\beta}_0)\| \leq \hm{G}_b(\hm{\beta}_0)\|\hm{\tilde{\beta}}_b-\hm{\beta}_0\|+M_1(\mathcal{D}_b)\|\hm{\tilde{\beta}}_b-\hm{\beta}_0 \|^2.
\end{equation}

Now we plug in~\eqref{pfeq:mean} into~\eqref{pfeq:beta0-beta}, and since all $\bm{\tilde{\beta}}_j=\bm{\beta}_0+O_p\left(\frac{1}{\sqrt{N_j}}\right)$ for $j=1,\dots,b-1$, we have
$
%\begin{split}
\bm{f}_b(\bm{\beta}_0)-\bm{f}_b(\bm{\tilde{\beta}}_b)
% = \left\{\bm{\tilde{G}}_b(\bm{\beta}_0) \right\}^\top
%\left\{\bm{\tilde{C}}_b(\bm{\beta}_0) \right\}^{-1} 
% \left\{\frac{1}{N_b}\bm{\tilde{G}}_b(\bm{{\beta}}_0)(\bm{\tilde{\beta}}_b-\bm{\beta}_0) +O_p\left(\frac{n_b}{N_b}\|\bm{\tilde{\beta}}_b-\bm{\beta}_0 \|^2\right)  \right\}
 =O_p\left(N_b^{-1/2}\right).
%\end{split}
$ By $i.i.d.$ assumption of $N_b$ samples, Law of Large Numbers and condition (C2), $\frac{1}{N_b}\bm{\tilde{G}}_b(\bm{\beta}_0)\overset{a.s.}{\to}\mathbb{G}(\bm{\beta}_0)$ and $\frac{1}{N_b}\bm{\tilde{C}}_b(\bm{\beta}_0)\overset{a.s.}{\to}\mathbb{C}(\bm{\beta}_0)$. Furthermore, by conditions~\ref{C3:bound_G} and~\ref{C5:positive_C}, $\mathbb{G}(\bm{\beta}_0)$ is of full rank and $\mathbb{C}^{-1}(\bm{\beta}_0)$ is positive-definite. Therefore, we have $\bm{\tilde{\beta}}_b\overset{p}{\to}\bm{\beta}_0$ as $N_b\to\infty$ which completes the induction proof. Furthermore, since all $\hm{\tilde{\beta}}_j$'s are $\sqrt{N_j}$-consistent for $j=1,\dots,b$, and according to equation~\eqref{pfeq:beta0-beta}, now we have
\[
\begin{split}
	\bm{f}_b(\hm{\beta}_0)-\bm{f}_b(\hm{\tilde{\beta}}_b)
	&=
	\left\{\frac{1}{N_b}\sum_{j=1}^{b}\hm{G}_j(\hm{{\beta}}_0)+O_p\left(\frac{1}{N_b}\sum_{j=1}^{b}\frac{n_j}{\sqrt{N_j}}\right) \right\}^\top\\ &\times\left\{\frac{1}{N_b}\sum_{j=1}^{b}\hm{C}_j(\hm{{\beta}}_0)+O_p\left(\frac{1}{N_b}\sum_{j=1}^{b}\frac{n_j}{\sqrt{N_j}}\right) \right\}^{-1}\\
	&\times\left\{\frac{1}{N_b}\sum_{j=1}^{b}\hm{G}_j(\bm{\beta}_0)+O_p\left(\frac{1}{N_b}\sum_{j=1}^{b}\frac{n_j}{\sqrt{N_j}}\right) \right\}(\hm{\tilde{\beta}}_b-\hm{\beta}_0) \\
	&=\mathbb{J}(\bm{\beta}_0)(\bm{\tilde{\beta}}_b-\bm{\beta}_0) +O_p(N_b^{-1}) \\
	&=O_p(N_b^{-1/2})
\end{split}
\]
where $\mathbb{J}(\hm{\beta}_0)= \mathbb{G}^\top(\hm{\beta}_0)\mathbb{C}^{-1}(\hm{\beta}_0)\mathbb{G}(\hm{\beta}_0)$.
Since both $\mathbb{G}(\bm{\beta})$ is of full rank and $\mathbb{C}(\bm{\beta})$ is positive-definite, it follows that $\mathbb{J}(\bm{\beta}_0)$ is also positive-definite, and the Central Limit Theorem implies
\begin{equation}\label{pfeq:asynormal}
\sqrt{N_b}(\hm{\tilde{\beta}}_b-\hm{\beta}_0) \overset{d}{\to} \mathcal{N}(\hm{0},\mathbb{J}^{-1}(\hm{\beta}_0)), \ \text{as}\ N_b\to\infty.
\end{equation}

%\subsection{Consistency in Scenario 1}\label{app:pf_QIF1_consist}

\subsection{Consistency in Scenario 1}\label{app:pf_QIF1_consist}
Since the initial estimator $\bm{\tilde{\beta}}_1$ satisfies
\[
\bG_1(\bm{\tilde{\beta}}_1)^\top \bC_1(\bm{\tilde{\beta}}_1)^{-1} \bg_1(\bm{\tilde{\beta}}_1) = \bm{0},
\]
and it follows that
\[
\bG_1(\bm{\tilde{\beta}}_1)^\top \bC_1(\bm{\tilde{\beta}}_1)^{-1} \bg_1(\bm{\tilde{\beta}}_1) - 
\bG_1(\bm{{\beta}}_0)^\top \bC_1(\bm{{\beta}}_0)^{-1} \bg_1(\bm{{\beta}}_0) = 
- \bG_1(\bm{{\beta}}_0)^\top \bC_1(\bm{{\beta}}_0)^{-1} \bg_1(\bm{{\beta}}_0).
\]
By condition~\ref{C3:bound_G}, $\bG_1(\bbeta)^\top\bC_1(\bbeta)^{-1}\bg_1(\bbeta)$ is continuously differentiable w.r.t. $\bbeta\in\Theta$. According to the mean value theorem for vector-valued functions, we have
\begin{equation}\label{eq:pf_first_batch}
\bm{\tilde{\beta}}_1 - \bbeta_0 = \left\{\int_{0}^{1} L(\bbeta_{m,1}) d\delta_1\right\}^{-1} \bG_1(\bbeta_0)^\top \bC_1(\bbeta_0)^{-1} \bg_1(\bbeta_0),
\end{equation}
where $L(\bbeta_{m,1}):= -\frac{\partial \left\{\bG_1(\bbeta)^\top \bC_1(\bbeta)^{-1}\bg_1(\bbeta)\right\}}{\partial \bbeta^\top}\mid_{\bbeta = \bbeta_{m,1}}$ is a positive-definite matrix by condition~\ref{C6:D1_pd}, and $\bbeta_{m,1} \in\mathbb{R}^p$ satisfying $\bbeta_{m,1} = \delta_1 \bm{\tilde{\beta}}_1 + (1 - \delta_1)\bbeta_0$ for some $\delta_1 \in (0, 1)$. Here, the integral of a matrix is component-wise.
%\end{comment}

Let $\mM_b =\left\{\sum_{j=1}^{b} \bG_j(\tbeta_j)\right\}^\top \left\{\sum_{j=1}^{b} \bC_j(\tbeta_j)\right\}^{-1}=\bm{\tilde{G}}_b^\top\bm{\tilde{C}}_b^{-1}$. Then for $b=2$, from equation (7), we have
\[
\begin{split}
\tbeta_2 - \bbeta_0 =& \tbeta_1 - \bbeta_0 + N_2^{-1} \bH_2^{-1} \mM_2\left\{\bm{\tilde{g}}_1(\tbeta_1)+\bg_2(\tbeta_2) \right\} \\
=& \tbeta_1 - \bbeta_0 + N_2^{-1}\bH_2^{-1}\mM_2\left\{\bg_1(\bbeta_0) + \bg_2(\bbeta_0)\right\}\\
& + N_2^{-1}\bH_2^{-1} \mM_2 \left\{\bar{\bG}_1(\bbeta_{m,1}) (\bbeta_0 - \tbeta_1)+\bar{\bG}_2(\bbeta_{m,2})(\bbeta_0 - \tbeta_2) \right\},
\end{split}
\]
where $\bar{\bG}_j(\bbeta_{m,j})=\int_{0}^{1}\bG_j(\bbeta_{m,j})d\delta_j$, and $\bbeta_{m,j}=\delta_j\tbeta_j + (1 - \delta_j)\bbeta_0$ for some $\delta_j\in(0, 1)$. Then it follows that
\[
\mA_{22}(\tbeta_2 - \bbeta_0) = \mA_{21}(\tbeta_1 - \bbeta_0) + N_2^{-1}\bH_2^{-1}\mM_2\left\{\bg_1(\bbeta_0) + \bg_2(\bbeta_0)\right\},
\]
where $\mA_{22} = \bI_p + N_2^{-1}\bH_2^{-1}\mM_2\bar{\bG}_2(\bbeta_{m,2})$ and $\mA_{21} = \bI_p - N_2^{-1}\bH_2^{-1}\mM_2\bar{\bG}_1(\bbeta_{m,1})$. 

Using the fact that $\bm{\tilde{g}}_b(\tbeta_b) = \bm{\tilde{g}}_{b-1}(\tbeta_{b-1})+\sum_{j=1}^{b-1}\bG_j(\tbeta_j)(\tbeta_{b-1} - \tbeta_b) + \bg_b(\tbeta_b)$, for an arbitrary $b$, we have
\[
\begin{split}
\tbeta_b - \bbeta_0 =& \tbeta_{b-1} - \bbeta_0 + N_b^{-1}\bH_b^{-1}\mM_b 
\left\{ 
\sum_{j=1}^{b}\bg_j(\tbeta_j) + \sum_{j=1}^{b-2} \bG_j(\tbeta_j)(\tbeta_{b-2} - \tbeta_{b-1}) \right.\\
&	\left. + \sum_{j=1}^{b-3} \bG_j(\tbeta_j)(\tbeta_{b-3} - \tbeta_{b-2}) + \dots + \bG_1(\tbeta_1)(\tbeta_1 - \tbeta_2) \right\}. \\
%	= & \tbeta_{b-1} - \bbeta_0 + N_b^{-1}\bH_b^{-1}\mM_b\left\{\sum_{j=1}^{b}\bg_j(\bbeta_0)\right\}
%	- N_b^{-1}\bH_b^{-1} \mM_b\left\{\sum_{j=1}^{b-2}\bG_j(\tbeta_j)\right\}(\tbeta_{b-1} - \bbeta_0) \\
%	& + N_b^{-1}\bH_b^{-1} \mM_b \left\{\sum_{j=1}^{b-2}\bG_j(\tbeta_j)(\tbeta_j - \bbeta_0) \right\} + N_b^{-1}\bH_b^{-1}\mM_b\left\{\sum_{j=1}^{b}\bG_j(\bbeta_{m,j})(\bbeta_0 - \tbeta_j) \right\} 
\end{split}
\]
After some simplifications, we obtain
\begin{equation}\label{eq:b_normality}
\tbeta_b - \bbeta_0	=  %\mA_{b,b}^{-1} \mA_{b,b-1}(\tbeta_{b-1} - \bbeta_0) +
\sum_{j=1}^{b-1} \mA_{b,b}^{-1} \mA_{b,j}(\tbeta_j - \bbeta_0) + \mA_{b,b}^{-1}N_b^{-1}\bH_b^{-1}\mM_b \left\{\sum_{j=1}^b\bg_j(\bbeta_0) \right\},
\end{equation}

where 
\[
\begin{aligned}
\mA_{b,b} &= \bI_p + N_b^{-1} \bH_b^{-1} \mM_b \bar{\bG}_b(\bbeta_{m,b}), \\
%\mA_{b,b-1} &= \bI_p - N_b^{-1}\bH_b^{-1} \mM_b \bar{\bG}_{b-1}(\bbeta_{m,b}) - N_b^{-1} \bH_b^{-1}\mM_b \left(\sum_{j=1}^{b-2}\bG_j(\tbeta_j)\right) \\
%& = N_b^{-1} \bH_b^{-1}\mM_b \left\{\bG_{b-1}(\tbeta_{b-1}) - \bG_{b-1}(\bbeta_{m,b-1}) \right\}, \\
\mA_{b,j} & = N_b^{-1} \bH_b^{-1} \mM_b \left\{\bG_j(\tbeta_j) - \bar{\bG}_j(\bbeta_{m,j}) \right\}, \ j \leq b - 1.
\end{aligned}
\]
Since $\bH_b$ is positive-definite by~\ref{C7:Hb_pd}, $\bG(\bbeta)$ is Lipschitz continuous in $\Theta$ by~\ref{C3:bound_G}, we have $\bH_b = O_p\left(\frac{N_{b-1}}{N_b} \right)$, $\mA_{b,b} =O_p\left(\frac{N_b}{N_{b-1}}\right)$, and $\mA_{b,j} = O_p\left(\frac{n_j}{N_{b-1}}\|\tbeta_j - \bbeta_0\| \right)$ for $j\leq b-1$. Then, we have $\|\tbeta_1 - \bbeta_0\|=O_p(1)$. First, we show that $\|\tbeta_b-\bbeta_0\|$ is bounded as $b\to\infty$. If $\|\tbeta_{j} - \bbeta_0 \|=O_p(1)$ for $j\leq b-1$, it is easy to see from equation~\eqref{eq:b_normality} that $\|\tbeta_b - \bbeta_0 \|\leq c N_b^{-1}\sum_{j=1}^{b-1}n_j\|\tbeta_{j} - \bbeta_0 \|^2 + O_p(N_b^{-1/2})= O_p(1)$, where $c$ is a positive constant that does not depend on $b$. By the induction theorem, for every $b$, $\|\tbeta_b - \bbeta_0 \| = O_p(1)$. 

Now we prove $\|\tbeta_b - \bbeta_0\|\to 0$ as $b\to\infty$ by contradiction. If this claim does not hold, there should exist a $\delta>0$ such that $\|\tbeta_b - \bbeta_0\|\geq \delta$ for all $b$. Then, it follows from $\|\tbeta_{j}- \bbeta_0\|=O_p(1)$ that $\|\tbeta_j - \bbeta_0\| = c_j\|\tbeta_{b-1} - \bbeta_0\|$ for some finite $c_j$ such that $c_j = O(\delta^{-1})$ for all $j$. According to equation~\eqref{eq:b_normality}, we have
\[
\begin{split}
\|\tbeta_b - \bbeta_0\| &\leq  \norm{\mA_{b,b}^{-1}\sum_{j=1}^{b-1}\mA_{b,j}(\tbeta_j - \bbeta_0)} + \norm{\mA_{b,b}^{-1}N_b^{-1}\bH_b^{-1}\mM_{b}\left(\sum_{j=1}^{b}\bg_j(\bbeta_0) \right)   }  \\
&\leq \left(\sum_{j=1}^{b-1}c_j\|\mA_{b,b}^{-1}\mA_{b,j}\| \right) \|\tbeta_{b-1} - \bbeta_0\| + O_p(N_b^{-1/2}) \\
& \leq c'\delta^{-2} \|\tbeta_{b-1} - \bbeta_0\|^2 + O_p(N_b^{-1/2}) \leq c'\delta^{-2} \|\tbeta_{b-1} - \bbeta_0 \|^2, 
\end{split}
\]
where $c'$ is a positive constant that does not depend on $b$, and the last inequality follows from $\|\tbeta_{b-1} - \bbeta_0\|\geq \delta > N_b^{-1/2}$ for large $b$. It follows that the above inequality shows that $\tbeta_b$ has a quadratic convergence rate, that is, $\|\tbeta_b - \bbeta_0\|\to 0$ as $b\to\infty$. Apparently, it contradicts to the claim. Thus, we finish the proof of consistency.
%({\color{red}Quadratic convergence requires that the coefficient is an asymptotic constant, and it is true under the assumption that there exists $\delta>0$ blabla... and $c_j$ is finite. However, the coefficient actually diverges, if we assume we know the order of $\|\tbeta_j-\bbeta_0\|=O_p(N_j^{-1/2})$, then $c_j=O_p(\sqrt{\frac{N_{b-1}}{N_j}})$. It follows that the coefficient in front of $\|\tbeta_{b-1}-\bbeta_0\|^2$ is $O(N_b^{-1}\sum_{j=1}^{b-1}n_jc_j^{2})=O(\sum_{j=1}^{b-1}\frac{n_j}{N_j})=O(\log b)$. ``Quadratic convergence" is under the assumption that $c_j$ is finite. }) 

\subsection{Asymptotic normality in Scenario 1}\label{app:pf_QIF1_normal}
According to equation~\eqref{eq:b_normality}, we have
\[
\begin{aligned}
\tbeta_b - \bbeta_0 & = \mA_{b,b}^{-1} \mA_{b,b-1} (\tbeta_{b-1} - \bbeta_0) + \sum_{j=1}^{b-2} \mA_{b,b}^{-1} \mA_{b,j}(\tbeta_j - \bbeta_0) + \mA_{b,b}^{-1} N_b^{-1} \bH_b^{-1} \mM_b \left\{\sum_{j=1}^b \bg_j(\bbeta_0)\right\}. \\
\end{aligned}
\]
Repeating this calculation recursively gives that
\[
\begin{aligned}
\tbeta_b - \bbeta_0 & = \mA_{b,b}^{-1} \left(\mA_{b,b-1}\mA_{b-1,b-1}^{-1} \mA_{b-1,b-2} + \mA_{b,b-2} \right) (\tbeta_{b-2} - \bbeta_0) \\
&\hspace{1cm} + \sum_{j=1}^{b-3} \mA_{b,b}^{-1} \left(\mA_{b,b-1}\mA_{b-1,b-1}^{-1} \mA_{b-1,j} + \mA_{b,j}\right)(\tbeta_j - \bbeta_0) + \mA_{b,b}^{-1} N_b^{-1} \bH_b^{-1} \mM_b \bg_b(\bbeta_0) \\
&\hspace{1cm} + \mA_{b,b}^{-1} \left(\mA_{b,b-1}\mA_{b-1,b-1}^{-1}N_{b-1}^{-1}\bH_{b-1}^{-1}\mM_{b-1}+N_{b}^{-1}\bH_b^{-1} \mM_b\right) \left\{\sum_{j=1}^{b-1} \bg_j(\bbeta_0) \right\}.
\end{aligned}
\]
Let $\tilde{\mA}_{b,j}^{(1)} = \mA_{b,b-1}\mA_{b-1,b-1}^{-1}\mA_{b-1,j}+\mA_{b,j}$ for $1\leq j \leq b-2$.
\[
\begin{split}
\tbeta_b - \bbeta_0 & = \mA_{b,b}^{-1} \left(\tilde{\mA}^{(1)}_{b,b-2} \mA_{b-2,b-2}^{-1} \mA_{b-2,b-3}+\tilde{\mA}^{(1)}_{b,b-3}\right)(\tbeta_{b-3} - \bbeta_0) \\
&\hspace{1cm} + \sum_{j=1}^{b-4} \mA_{b,b}^{-1} \left(\tilde{\mA}^{(1)}_{b,b-2}\mA_{b-2,b-2}^{-1}\mA_{b-2,j}+\tilde{\mA}_{b,j}^{(1)} \right)(\tbeta_j - \bbeta_0) + \mA_{b,b}^{-1} N_b^{-1}\bH_b^{-1}\mM_b\bg_b(\bbeta_0) \\
& \hspace{1cm} + \mA_{b,b}^{-1} \left(\mA_{b,b-1}\mA_{b-1,b-1}^{-1} N_{b-1}^{-1}\bH_{b-1}^{-1}\mM_{b-1}+N_b^{-1}\bH_b^{-1}\mM_b\right) \bg_{b-1}(\bbeta_0) \\
&\hspace{1cm} + \mA_{b,b}^{-1}\left(\tilde{\mA}_{b,b-2}^{(1)}\mA_{b-2,b-2}^{-1} N_{b-2}^{-1} \bH_{b-2}^{-1}\mM_{b-2}+\mA_{b,b-1}\mA_{b-1,b-1}^{-1} N_{b-1}^{-1}\bH_{b-1}^{-1}\mM_{b-1}\right. \\
&\left. \hspace{2cm}+ N_b^{-1}\bH_b^{-1}\mM_b\right) \left\{\sum_{j=1}^{b-2}\bg_j(\bbeta_0) \right\}...
\end{split}
\]

Let $\tilde{\mA}_{b,j}^{(2)} = \tilde{\mA}_{b,b-2}^{(1)}\mA_{b-2,b-2}^{-1}\mA_{b-2,j} + \tilde{\mA}_{b,j}^{(1)}$ for $1\leq j \leq b-3$. Then, it follows that
\[
\begin{split}
\tbeta_b - \bbeta_0 = & \mA_{b,b}^{-1} \left(\tilde{\mA}_{b,b-3}^{(2)}\mA_{b-3,b-3}^{-1}\mA_{b-3,b-4} +\tilde{A}_{b,b-4}^{(2)} \right) (\tbeta_{b-4} - \bbeta_0) \\
& + \sum_{j=1}^{b-5} \mA_{b,b}^{-1} \left(\tilde{\mA}_{b,b-3}^{(2)} \mA_{b-3,b-3}\mA_{b-3,j} + \tilde{\mA}_{b,j}^{(2)}\right) (\tbeta_j - \bbeta_0) 
+ \mA_{b,b}^{-1} N_b^{-1}\bH_b^{-1} \mM_b\bg_b(\bbeta_0) \\
& + \mA_{b,b}^{-1} \left(\mA_{b,b-1}\mA_{b-1,b-1}^{-1} N_{b-1}^{-1} \bH_{b-1}^{-1} \mM_{b-1} + N_b^{-1}\bH_b^{-1}\mM_b \right) \bg_{b-1}(\bbeta_0) \\
& + \mA_{b,b}^{-1} \left(\tilde{\mA}_{b,b-2}^{(1)} \mA_{b-2,b-2}^{-1} N_{b-2}^{-1}\bH_{b-2}^{-1} \mM_{b-2} \right. \\
&\left.
\hspace{1cm}+\mA_{b,b-1} \mA_{b-1,b-1}^{-1} N_{b-1}^{-1}\bH_{b-1}^{-1} \mM_{b-1} + N_b^{-1}\bH_b^{-1}\mM_b
\right) \bg_{b-2}(\bbeta_0) \\
& + \mA_{b,b}^{-1} \left(\tilde{\mA}_{b,b-3}^{(2)}\mA_{b-3,b-3}^{-1} N_{b-3}^{-1}\bH_{b-3}^{-1} \mM_{b-3} + \tilde{\mA}_{b,b-2}^{(1)}\mA_{b-2,b-2}^{-1} N_{b-2}^{-1}\bH_{b-2}^{-1} \mM_{b-2}\right.\\
&\left. \hspace{1cm} + \mA_{b,b-1}\mA_{b-1,b-1}^{-1} N_{b-1}^{-1}\bH_{b-1}^{-1}\mM_{b-1} + N_{b}^{-1}\bH_b^{-1} \mM_b
\right) \left\{\sum_{j=1}^{b-3} \bg_j(\bbeta_0) \right\}.
\end{split}
\]

Finally, denote $\tilde{\mA}_{b,b-1}^{(0)}:=\mA_{b,b-1}$ and $\tilde{\mA}_{b,b}^{(-1)} := \mA_{b,b}$, then we obtain that 
\begin{equation}\label{eq:pf_normality}
\begin{split}
\tbeta_b - \bbeta_0
%& \mA_{b,b}^{-1}\tilde{\mA}_{b,2}^{(b-3)}(\tbeta_2 - \bbeta_0) + \mA_{b,b}^{-1}\tilde{\mA}_{b,1}^{(b-3)}(\tbeta_1 - \bbeta_0) \\
%& + \mA_{b,b}^{-1}\sum_{j=3}^{b} \left\{\left(\sum_{j' = j}^{b}\tilde{\mA}_{b,j'}^{(b - j'-1)}\mA_{j',j'}^{-1}N_{j'}^{-1}\bH_{j'}^{-1}\mM_{j'}\right)\bg_j(\bbeta_0) \right\} \\
%& + \mA_{b,b}^{-1} \sum_{j=1}^{2} \left\{\left(\sum_{j'=3}^{b}\tilde{\mA}_{b,j'}^{(b - j'-1)}\mA_{j',j'}^{-1}N_{j'}^{-1}\bH_{j'}^{-1}\mM_{j'} \right)\bg_j(\bbeta_0) \right\} \\
= & \mA_{b,b}^{-1} \left(\tilde{\mA}_{b,2}^{(b-3)}\mA_{22}^{-1}\mA_{21}+\tilde{\mA}_{b,1}^{(b-3)}\right)(\tbeta_1 - \bbeta_0) \\
& + \mA_{b,b}^{-1}\sum_{j=2}^{b} \left\{\left(\sum_{j' = j}^b\tilde{\mA}_{b,j'}^{(b-j'-1)}\mA_{j',j'}^{-1}N_{j'}^{-1}\bH_{j'}^{-1}\mM_{j'}\right)\bg_j(\bbeta_0) \right\} \\
& + \mA_{b,b}^{-1} \left\{\left(\sum_{j' = 2}^b \tilde{\mA}_{b,j'}^{(b-j'-1)}\mA_{j',j'}^{-1} N_{j'}^{-1}\bH_{j'}^{-1}\mM_{j'}\right)\bg_1(\bbeta_0) \right\}.
\end{split}
\end{equation}
Since $\bH_b=O_p\left(\frac{N_{b-1}}{N_b}\right)$, $\mA_{b,b}=O_p\left(\frac{N_b}{N_{b-1}} \right)$, and $\mA_{b,j}=O_p\left(\frac{n_j}{N_{b-1}}\|\tbeta_j - \bbeta_0\| \right)$ for $j\leq b - 1$,
\[
\begin{aligned}
\tilde{\mA}_{b,b-j}^{(1)} = & \mA_{b,b-1}\mA_{b-1,b-1}^{-1}\mA_{b-1,b-j} + \mA_{b,b-j} \\
= & O\left\{\frac{n_{b-j}\|\tbeta_{b-j} - \bbeta_0\|}{N_{b-1}}\left(1 + \frac{n_{b-1}\|\tbeta_{b-1}-\bbeta_0\|}{N_{b-1}}\right) \right\}, \ j \geq 2, \\
\tilde{\mA}_{b,b-j}^{(2)} = & \tilde{\mA}_{b,b-2}^{(1)} \mA_{b-2,b-2}^{-1}\mA_{b-2,b-j} + \tilde{\mA}_{b,b-j}^{(1)} \\
= & O\left\{\frac{n_{b-j}\|\tbeta_{b-j} - \bbeta_0\|}{N_{b-1}}\left(1+\frac{n_{b-1}\|\tbeta_{b-1}-\bbeta_0\|}{N_{b-1}}\right)\left(1+\frac{n_{b-2}\|\tbeta_{b-2}-\bbeta_0\|}{N_{b-2}}\right)\right\}, \ j \geq 3, \\
%\tilde{\mA}_{b,b-j}^{(3)}  = & \tilde{\mA}_{b,b-3}^{(2)}\mA_{b-3,b-3}^{-1}\mA_{b-3,b-j} + \tilde{\mA}_{b,b-j}^{(2)} \\
%= &O\left\{\frac{n_{b-j}\|\tbeta_{b-j} - \bbeta_0\|}{N_{b-1}}\prod_{k=1}^3\left(1+\frac{n_{b-k}\|\tbeta_{b-k}-\bbeta_0\|}{N_{b-k}}\right)\right\}, \ j \geq 4, \\
&\vdots\\
\tilde{\mA}_{b,j'}^{(b-j-1)} = &\tilde{\mA}_{b,j+1}^{(b-j-2)}\mA_{j+1,j+1}^{-1}\mA_{j+1,j'} + \tilde{\mA}_{b,j'}^{(b-j-2)} \\
=& O\left\{\frac{n_{j'}\|\tbeta_{j'} - \bbeta_0\|}{N_{b-1}}\prod_{k=1}^{b-j-1}\left(1 + \frac{n_{b-k}\|\tbeta_{b-k} - \bbeta_0\|}{N_{b-k}}\right)\right\}, \ j'\leq j.
\end{aligned}
\]

Let $a_j = \sum_{j'=j}^b\tilde{\mA}_{b,j'}^{(b - j' -1)}\mA_{j',j'}^{-1} N_{j'}^{-1}\bH_{j'}^{-1}\mM_{j'}$. After some calculations, we have
\[
\begin{aligned}
a_b & = \tilde{\mA}_{b,b}^{(-1)} \mA_{b,b}^{-1}N_b^{-1}\bH_b^{-1}\mM_b = O\left(\frac{1}{N_{b-1}}\right), \\
a_{b-1} & = \tilde{\mA}_{b,b-1}^{(0)} \mA_{b-1,b-1}^{-1} N_{b-1}^{-1} \bH_{b-1}^{-1}\mM_{b-1} + a_b 
%& = O\left(\frac{\|\tbeta_{b-1} - \bbeta_0\|}{b-1}\frac{b-2}{b-1}\frac{b-1}{b-2}\frac{1}{b-1} + \frac{1}{b-1}\right) 
= O\left\{\frac{1}{N_{b-1}}\left(1+\frac{n_{b-1}\|\tbeta_{b-1} - \bbeta_0\|}{N_{b-1}} \right) \right\}, \\
%a_{b-2} & = \tilde{\mA}_{b,b-2}^{(1)} \mA_{b-2,b-2}^{-1} N_{b-2}^{-1}\bH_{b-2}^{-1} \mM_{b-2} + a_{b-1} \\
%&=O\left\{\frac{1}{N_{b-1}}\left(1+\frac{n_{b-1}\|\tbeta_{b-1}-\bbeta_0\|}{N_{b-1}}\right)\left(1+\frac{n_{b-2}\|\tbeta_{b-2} - \bbeta_0\|}{N_{b-2}}\right)\right\}, \\
&\vdots \\
a_j & = O\left\{\frac{1}{N_{b-1}}\prod_{k=1}^{b-j}\left(1 + \frac{n_{b-k}\|\tbeta_{b-k} - \bbeta_0\|}{N_{b-k}}\right)  \right\}.
\end{aligned}
\]
Combining the above terms with equations~\eqref{eq:pf_first_batch} and~\eqref{eq:pf_normality} leads to 
\begin{equation}\label{eq:b_normality_simple}
\tbeta_b - \bbeta_0 = \mA_{b,b}^{-1} \left[\tilde{\mA}_{b,1}^{(b-2)}\left\{\int_{0}^{1} L(\bbeta_{m,1})d\delta_1\right\} ^{-1}\bG_1(\bbeta_0)^\top\bC_1(\bbeta_0)^{-1} + a_2\right]\bg_1(\bbeta_0) + \mA_{b,b}^{-1}\sum_{j=2}^{b} a_j\bg_j(\bbeta_0).
\end{equation}

Since we have shown the consistency of $\tbeta_b$ as $b\to\infty$, we further assume $\|\tbeta_b - \bbeta_0\| = O_p(N_b^{-\tau})$ for some $\tau>0$. We next show $\tau = 0.5$. Note that for positive series $\{s_k \}_{k=1}^b$, $\prod_{k=1}^b(1 + s_k)$ converges if and only if $\sum_{k=1}^{b}s_k$ converges. For $\tau>0$, and uniformly bounded $n_{b-k}$'s by condition~\ref{C8:finite_nb}, the over-harmonic series
$\sum_{k=1}^{b-1} n_{b-k}N_{b-k}^{-(1+\tau)}$ converges, and it follows that $\prod_{k=1}^{b-1}\left\{1+n_{b-k}N_{b-k}^{-(1+\tau)} \right\}$ converges. Then we have $a_1 = O\left(N_{b-1}^{-1}\right)$. Similarly, we have $a_2=O\left(N_{b-2}^{-1}\right)$. Then it follows from equation~\eqref{eq:b_normality_simple} that
\[
\|\tbeta_b - \bbeta_0 \| = O_p\left\{\frac{\sqrt{n_1}}{N_{b}} + \frac{N_{b-1}}{N_b} \left(\sum_{j=2}^{b}n_j a_j^2 \right)^{1/2}\right\}.
\]
Note that for any $x\geq 0$, $\log (1+x)\leq x$. Thus, we have
\[
\sum_{k=1}^{b-1} \log\left(1+n_{b-k}N_{b-k}^{-(1+\tau)}\right) \leq \sum_{k=1}^{b-1} n_{b-k}N_{b - k}^{-(1+\tau)}.
\]
Then, it follows that 
\[
\begin{split}
\sum_{j=2}^{b} n_j a_j^2 & \leq \frac{1}{N_{b-1}^2} \sum_{j=2}^{b}n_j \exp \left\{\left(\sum_{k=1}^{b-j}n_{b-k}N_{b-k}^{-(1+\tau)}\right)^2 \right\} \\
& \leq \frac{1}{N_{b-1}^2} \sum_{j=2}^{b}n_j\exp \left\{\left(\sum_{k=1}^{b-2}n_{b-k}N_{b-k}^{-(1+\tau)} \right)^2\right\} 
%&\leq \frac{1}{N_{b-1}^2} \sum_{j=2}^{b}n_j \exp \left\{ \left(\log \frac{N_{b-1}}{n_1}\right)^2 \right\} \\
= O_p\left(\frac{N_b - n_1}{N_{b-1}^2}\right)=O_p(N_b^{-1}).
\end{split}
\]
Thus, we have $\|\tbeta_b -\bbeta_0 \| = O_p(N_b^{-1/2})$. Finally, using the expression~\eqref{eq:b_normality} and equation~\eqref{eq:lemma11} in Lemma~\ref{pf:lemma}, we have 
\[
\begin{split}
\tbeta_b - \bbeta_0 & = \mA_{b,b}^{-1} N_b^{-1}\bH_b^{-1} \mM_b \left\{\sum_{j=1}^{b}\bg_j(\bbeta_0) \right\} + O_p\left(\norm{\sum_{j=2}^{b-2}\mA_{b,b}^{-1}\mA_{b,j}(\tbeta_j -\bbeta_0) } \right) \\
& = \mA_{b,b}^{-1}N_b^{-1}\bH_b^{-1}\mM_b\left\{\sum_{j=1}^{b}\bg_j(\bbeta_0)\right\}
+ O_p\left(\frac{1}{N_b}\sum_{j=2}^{b-2}n_j\|\tbeta_j - \bbeta_0\|^2\right) \\
& = \mA_{b,b}^{-1} N_b^{-1}\bH_b^{-1}\mM_b\left\{\sum_{j=1}^{b}\bg_j(\bbeta_0)\right\}
+O_p\left(\frac{1}{N_b}\sum_{j=2}^{b-2}\frac{n_j}{N_j}\right) \\
& = \mA_{b,b}^{-1}N_b^{-1} \bH_b^{-1} \mM_b\left\{\sum_{j=1}^{b}\bg_j(\bbeta_0) \right\} + O_p\left(\frac{1}{N_b}\log \frac{N_{b-2}}{n_1} \right).
\end{split}
\]
By the weak law of large numbers, the asymptotic covariance is
\[
\begin{split}
N_b^{-1}\bH_b^{-1}\mM_b\left\{\sum_{j=1}^{b}\bC_j(\bbeta_0)\right\} \mM_b^\top(\bH_b^{-1})^\top
&=(\mathbb{G}^\top\mathbb{C}^{-1}\mathbb{G})^{-1}\mathbb{G}^\top\mathbb{C}^{-1}\mathbb{C}\mathbb{C}^{-1}\mathbb{G} (\mathbb{G}^\top\mathbb{C}^{-1}\mathbb{G})^{-1} +o_p(1)\\
&=(\mathbb{G}^\top\mathbb{C}^{-1}\mathbb{G})^{-1}+o_p(1).
\end{split}
\]
It then follows from the Slutsky's Theorem and the Central Limit Theorem, 
\[
\sqrt{N_b}(\tbeta_b - \bbeta_0) = \left\{\frac{1}{N_b}\sum_{j=1}^{b}\bG_j(\bbeta_0)\right\}^{-1} \left\{\frac{1}{\sqrt{N_b}}\sum_{j=1}^{b}\bg_j(\bbeta_0)\right\} + O_p\left(\frac{\log b}{\sqrt{N_b}}\right) 
\overset{d}{\to} \mathcal{N}_p(\bm{0}, \mathbb{J}^{-1}(\bbeta_0)),
\]
where $\mathbb{J}(\bbeta_0)=\mathbb{G}^\top(\bbeta_0)\mathbb{C}^{-1}(\bbeta_0)\mathbb{G}(\bbeta_0)$ is the Godambe information matrix.

%\section{Error bound}
%Since we have shown that $\|\tbeta_b-\bbeta_0\|=O_p(N_b^{-1/2})$, and combing it with equation~\eqref{eq:b_normality}, we have
%\[
%\begin{split}
%\mathbb{E}\|\tbeta_b-\bbeta_0\|&\leq  \mathbb{E}\norm{\mA_{b,b}^{-1}\sum_{j=1}^{b-1}\mA_{b,j}(\tbeta_j - \bbeta_0)} + \mathbb{E}\norm{\mA_{b,b}^{-1}N_b^{-1}\bH_b^{-1}\mM_{b}\left(\sum_{j=1}^{b}\bg_j(\bbeta_0) \right)   }  \\
%&\leq C_0 \frac{N_{b-1}}{N_b}\sum_{j=1}^{b-1}\frac{n_j}{N_{b-1}}\mathbb{E}\|\tbeta_j-\bbeta_0\|^2 + \frac{C_2}{\sqrt{N_b}} \\
%& \leq \frac{C_1}{N_b} \sum_{j=1}^{b-1}\frac{n_j}{N_j} +\frac{C_2}{\sqrt{N_b}} \leq \frac{C_1}{N_b}\left(1+\log \frac{N_{b-1}}{n_1}\right) + \frac{C_2}{\sqrt{N_b}},
%\end{split}
%\]
%where the last inequality follows from equation~\eqref{eq:lemma11} in Lemma~\ref{pf:lemma}. $C_1$ and $C_2$ are two positive constants that do not depend on $N_b$. Furthermore, since $\lim_{b\to\infty}\frac{\log N_{b}}{N_{b}}=0$, this upper bound vanishes fast to 0 rather than accumulates over the recursive procedure as $b$ increases.

%\section*{Supplementary Material}\label{sec:SM}
%\begin{description}

%\item[Supplementary material:] 
%{\color{red}This file includes one table and one figure from simulation studies and an additional table from real data analysis. It also includes a section ``Renewable GEE" with derivation of renewable estimation and incremental inference method in the generalized estimating equations. (PDF)}

%\item[Renewable GEE:] Derivation of renewable estimation and inference method in the generalized estimating equations. (PDF)

%\end{description}

\bigskip
\begin{center}
	{\large\bf SUPPLEMENTARY MATERIAL}
\end{center}

\section{Tables and Figures}
\begin{table}[h]
	\caption{\label{tab:empirical}Empirical type I error rate ($\times 10^{-3}$) under a total number of $B=100$ data batches with different data batch size $n_b$ and various significance level $\alpha$. In the calculation of empirical power, the locations of two contaminated data batches are $\tau_1=25$ and $\tau_2=75$. Results are summarized over $500$ replications. }
	\begin{center}
		\begin{tabular} {l l l l l | l l l l | l l l l}
			\hline
			\hline
			&	\multicolumn{4}{c|}{\makecell{Empirical type I error\\ rate ($\times 10^{-3}$)}} 	
			&   \multicolumn{8}{c}{Empirical power (\%)} \\
			&	\multicolumn{4}{c|}{$d=0$} 	
			&   \multicolumn{4}{c}{$d=0.5$}
			&   \multicolumn{4}{c}{$d=1.0$} \\
			\hline
			&\multicolumn{4}{c|}{$n_b$} 	&\multicolumn{4}{c|}{$n_b$}  &\multicolumn{4}{c}{$n_b$}\\
			$\alpha\times 10^{-3}$  &50  &100   &200  &400
			&50  &100   &200  &400
			&50  &100   &200  &400 \\
			\hline
			100      
			&106.4  &106.3  &102.9  &104.7
			&54.8  &87.4  &99.8  &100
			&97.2  &100  &100  &100 \\
			50        
			&49.6 &51.5 &49.5  &53.6
			&40.4  &78.6  &99.0  &100
			&94.6  &100  &100  &100 \\
			10        
			&6.8  &9.0  &9.7  &12.5
			&13.8  &55.4  &96.2  &100
			&77.0  &100  &100  &100 \\
			1           
			&0.3   &0.5  &1.5  &1.6
			&24.0  &25.4  &86.0  &100
			&42.6  &98.4  &100  &100 \\
			0.005  
			&0  &0      &0  &0 
			&0       &20.0   &42.4  &99.4
			&22.0  &77.4  &100  &100 \\
			\hline
		\end{tabular}
	\end{center}
\end{table}

\begin{figure}[h]
	\centering
	\includegraphics[width=\linewidth]{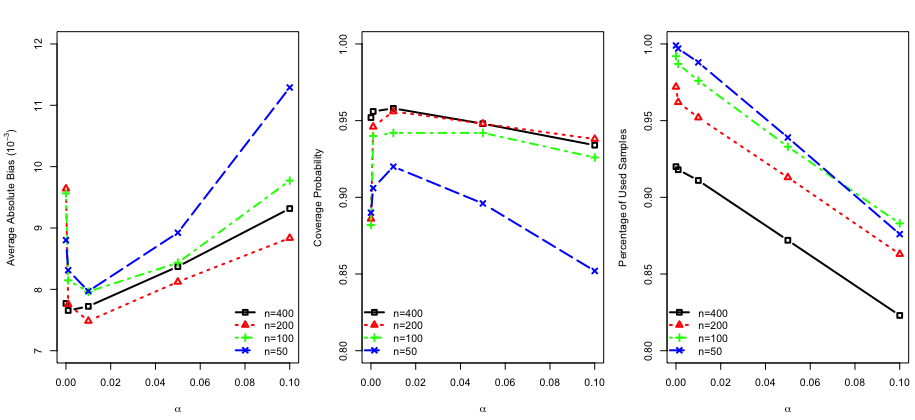}
	\caption{For fixed $N_B=10^4$, the relationship between type I error, estimation bias, coverage probability and percentage of used data.}
	\label{fig:bias_cp_N}
\end{figure}

\begin{table}[h]
	\caption{\label{tab:CDS_compare_QIF}Results from the offline QIF method ($N_B=18,832$), the proposed RenewQIF in logistic model with data batch 8 ($N_B=18,832$, $B=28$), and RenewQIF$_{qc}$ without data batch 8 ($N_0=18,157$, $B=27$). }
	\begin{center}
		\begin{tabular}{lccc|ccc|ccc}	
			\hline
			\hline
			&\multicolumn{3}{c|}{QIF} & \multicolumn{3}{c|}{RenewQIF} & \multicolumn{3}{c}{RenewQIF$_{qc}$}\\	
			&\multicolumn{1}{c}{$\hat{\beta}^\star_B$} 
			&\multicolumn{1}{c}{ASE}
			&\multicolumn{1}{c|}{$p$} 
			&\multicolumn{1}{c}{$\tilde{\beta}_B$} 
			&\multicolumn{1}{c}{ASE}
			&\multicolumn{1}{c|}{$p$} 
			&\multicolumn{1}{c}{$\tilde{\beta}_B$} 
			&\multicolumn{1}{c}{ASE}
			&\multicolumn{1}{c}{$p$}\\	
			\hline
			Intercept  &-0.90  &0.088 &0.000
			&-0.89  &0.092 &0.000
			&-0.91  &0.094 &0.000
			\\
			Young     &-0.22   &0.050  &0.000
			&-0.22 &0.051 &0.000
			&-0.22 &0.052 &0.000
			\\
			Old     &0.57   &0.061  &0.000
			&0.57 &0.062 &0.000
			&0.57 &0.063 &0.000
			\\
			Seat Belt  &-1.16  &0.046 &0.000
			&\bf-1.15  &0.047 &0.000
			&\bf-1.14  &0.048 &0.000
			\\
			Drinking &0.36  &0.049 &0.000
			&\bf0.36 &0.049 &0.000
			&\bf0.37 &0.050 &0.000
			\\
			Speed Limit &0.17 &0.034 &0.000
			&0.17 &0.035 &0.000
			&0.17 &0.036 &0.000
			\\
			Vehicle Weight &-0.024 &0.028 &0.399
			&\bf-0.026  &0.029 &0.365
			&\bf-0.025  &0.029 &0.385 \\
			
			Air Bag &-1.00  &0.050 &0.000
			&-1.00 &0.052 &0.000
			&-1.00 &0.053 &0.000\\
			
			Number of Lanes &-0.10 &0.035 &0.005
			&\bf-0.10 &0.037 &0.008
			&\bf-0.11  &0.037  &0.003 \\
			
			Drug Use &0.96  &0.044 &0.000
			&\bf0.95  &0.045 &0.000
			&\bf0.94 &0.046 &0.000\\
			
			Distraction &-0.45 &0.035 &0.000
			&-0.45 &0.036 &0.000
			&-0.45 &0.036  &0.000\\
			
			Surface Condition &0.16 &0.046 &0.001
			&0.15 &0.049 &0.002
			&0.16  &0.050  &0.002\\
			
			Previous Accident &1.17 &0.059 &0.000
			&1.16   &0.061  &0.000
			&1.16   &0.062  &0.000 \\
			\hline
		\end{tabular}
	\end{center}
\end{table}

\clearpage
\section{Renewable GEE}
At an arbitrary time point $b$, let $\hm{\psi}_b(\mathcal{D}_b;\hm{\beta},\hm{\alpha})=\sum_{i\in \mathcal{D}_b}\hm{D}_{i}^\top\hm{\Sigma}_{i}^{-1}(\hm{y}_{i}-\hm{\mu}_{i})$ be the estimating function for data batch $\mathcal{D}_b$, and the corresponding negative gradient and sample variance matrices are denoted by $\hm{S}_b(\mathcal{D}_b;\hm{\beta},\hm{\alpha})=\sum_{i\in \mathcal{D}_b}\hm{D}_{i}^\top \hm{\Sigma}_{i}^{-1} \hm{D}_{i}$ and $\hm{V}_b(\mathcal{D}_b;\hm{\beta},\hm{\alpha})=\sum_{i\in \mathcal{D}_b}\hm{D}_{i}^\top \hm{\Sigma}_{i}^{-1} (\hm{y}_{i}-\hm{\mu}_{i}) (\hm{y}_{i}-\hm{\mu}_{i})^\top \hm{\Sigma}_{i}^{-1} \hm{D}_{i}$, respectively. 

Similar to the derivation of RenewQIF, we propose a renewable GEE (RenewGEE) estimator $\hm{\tilde{\beta}}_b$ of $\hm{\beta}_0$ as a solution to the following incremental estimating equation:
\begin{equation}\label{app:GEE_b}
\sum_{j=1}^{b-1}\hm{S}_j(\mathcal{D}_j;\hm{\tilde{\beta}}_{j})(\hm{\tilde{\beta}}_{b-1}-\hm{\tilde{\beta}}_b)+\hm{\psi}_b(\mathcal{D}_b;\hm{\tilde{\beta}}_b)=\hm{0}.
\end{equation}
Let $\hm{\tilde{S}}_b=\sum_{j=1}^{b}\hm{S}_j(\mathcal{D}_j;\hm{\tilde{\beta}}_j)$ denote the aggregated negative gradient matrix. Solving equation~\eqref{app:GEE_b} may be easily done by the following incremental updating algorithm:
\begin{equation}\label{app:GEE_b_algorithm}
\hm{\tilde{\beta}}_b^{(r+1)}=\hm{\tilde{\beta}}_b^{(r)}+\left\{\hm{\tilde{S}}_{b-1}+\hm{S}_b(\mathcal{D}_b;\hm{\tilde{\beta}}_{b}^{(r)}) \right\}^{-1} \hm{\tilde{\psi}}_b^{(r)},
\end{equation}
where $\hm{\tilde{\psi}}_b^{(r)}=\hm{\tilde{S}}_{b-1}(\hm{\tilde{\beta}}_{b-1}-\hm{\tilde{\beta}}_b^{(r)})+\hm{\psi}_b(\mathcal{D}_b;\hm{\tilde{\beta}}_b^{(r)})$ denotes the adjusted estimating function. In equation~\eqref{app:GEE_b_algorithm}, both the gradient and the adjusted estimating function use the subject-level data of current data batch $\mathcal{D}_b$ and summary statistics $\left\{\hm{\tilde{\beta}}_{b-1},\hm{\tilde{S}}_{b-1},\hm{\tilde{\alpha}}_{b-1},\tilde{\phi}_{b-1} \right\}$ from historical data, where consistent estimators of nuisance parameters are updated recursively according to
$
\hm{\tilde{\alpha}}_b=\tilde{w}_{b-1}\hm{\tilde{\alpha}}_{b-1}+w_b\hm{\hat{\alpha}}_b(\hm{\tilde{\beta}}_b,\hat{\phi}_b)$ and
$\tilde{\phi}_b=\tilde{w}_{b-1}\tilde{\phi}_{b-1}+w_b\hat{\phi}_b$,
where $\hm{\hat{\alpha}}_b$ and $\hat{\phi}_b$ are the estimated nuisance parameters using only $\mathcal{D}_b$, and the weights are $\tilde{w}_{b-1}=\frac{mN_{b-1}-p}{mN_b-p}$ and $w_b=\frac{m-p}{mN_b-p}$. In addition, for estimating the sandwich estimator, we also need the aggregated sample variance matrix for $\hm{\tilde{\psi}}_b$, denoted by $\hm{\tilde{V}}_b=\sum_{j=1}^{b}\hm{V}_j(\mathcal{D}_j;\hm{\tilde{\beta}}_j)$. Then we calculated the estimated asymptotic covariance matrix $\widetilde{\hm{\Sigma}_b}(\hm{\beta}_0)=N_b\left\{\hm{\tilde{S}}_b^\top\hm{\tilde{V}}_b^{-1}\hm{\tilde{S}}_b \right\}^{-1}$.
It follows that the estimated asymptotic variance matrix for the renewable GEE $\hm{\tilde{\beta}}_b$ is given by
$
\text{var}(\hm{\tilde{\beta}}_b):=\left\{\hm{\tilde{S}}_b^\top\hm{\tilde{V}}_b^{-1}\hm{\tilde{S}}_b \right\}^{-1}.
$

\bibliographystyle{authordate1}
\bibliography{paper-ref}

\end{document}